\documentclass[prd,article,nofootinbib
,twocolumn,preprintnumbers,showpacs]{revtex4}

\usepackage{graphicx,amsfonts,color,comment,amsmath,hyperref}
\usepackage{times}
\usepackage{amssymb}	
\newcommand{\be}{\begin{equation}}
\newcommand{\ee}{\end{equation}}
\newcommand{\bea}{\begin{eqnarray}}
\newcommand{\eea}{\end{eqnarray}}
\newcommand{\mjd}{{\sc Majorana Demonstrator}}

\begin{document}

\title{Integrating In Dark Matter Astrophysics at Direct Detection Experiments}

\author{Alexander Friedland}
\email{friedland@lanl.gov}
\author{Ian M. Shoemaker}
\email{ianshoe@lanl.gov}
\affiliation{Theoretical Division T-2, MS B285, Los Alamos National Laboratory, Los Alamos, NM 87545, USA}

\date{May 29, 2013}

\begin{abstract}

We study the capabilities of the \mjd, a neutrinoless double-beta decay experiment currently under construction at the Sanford Underground Laboratory, as a light WIMP detector. For a cross section near the current experimental bound, the \mjd~ should collect hundreds or even thousands of recoil events. This opens up the possibility of simultaneously determining the physical properties of the dark matter and its local velocity distribution, directly from the data. We analyze this possibility and find that allowing the dark matter velocity distribution to float considerably worsens the WIMP mass determination. This result is traced to a previously unexplored degeneracy between the WIMP mass and the velocity dispersion. We simulate spectra using both isothermal and Via Lactea II velocity distributions and comment on the possible impact of streams. We conclude that knowledge of the dark matter velocity distribution will greatly facilitate the mass and cross section determination for a light WIMP.

 \end{abstract}
\preprint{LA-UR-12-26557}
\maketitle

\section{Introduction}

Direct detection experiments offer the possibility of a non-gravitational detection of dark matter (DM), the most common form of matter in the Universe~\cite{Goodman:1984dc}.  Swift progress is being made in this field, with the LUX and XENON1T experiments aiming to push the bound on the spin-independent cross section of a 100 GeV WIMP all the way to the  $10^{-47}$ cm$^{-2}$ level. A no less dramatic development may occur on the light  WIMP front ($\lesssim$ 10 GeV), where the improvement relative to the current bound may be even greater, as a result of the \mjd~\cite{MajThesis,Finnerty:2012dr} (see Fig.~\ref{sensitivity} below). 
If the light WIMP cross section lies near its current bound, the \mjd~ is poised to collect hundreds or even thousands of recoil events. This would open up an interesting possibility of not only discovering the DM particle, but also accurately measuring its properties. Here, we explore this possibility and the issues that arise in connection with it.

Light WIMPs are not without theoretical or experimental motivation. For example, on the theoretical side, 1-10 GeV DM is typical in models relating the baryon and DM abundances~\cite{Nussinov:1985xr,Barr:1990ca,Hooper:2004dc,Agashe:2004bm,Farrar:2005zd,Gudnason:2006ug,Kaplan:2009ag} (see \cite{Tulin:2012re} and references therein for recent work).  
In these models, the DM has a nonzero particle-antiparticle asymmetry, thereby suppressing the indirect signatures of DM annihilation \footnote{See, however, \cite{Graesser:2011wi} for a counterexample.}.  Establishing such low-mass DM at a collider will be difficult since although the Tevatron and the LHC can detect the missing energy associated with light DM, they lose sensitivity to the mass of DM when it is much less than 100 GeV (see e.g.,~\cite{Goodman:2010yf}). The missing energy could then be attributed to a variety of sources, from novel neutrino interactions \cite{Friedland:2011za} to extra-dimensional particles\cite{ArkaniHamed:1998rs,Abulencia:2006kk}. Hence, low-threshold direct detection experiments such as the \mjd~ may offer the best way to test this well-motivated class of models.

On the experimental side, the interest of the community has been piqued 
by the CoGeNT and DAMA experiments, which have claimed signs of light dark matter ~\cite{Bernabei:2008yi,Aalseth:2010vx}. Most recent hints in this mass window come from the CDMS experiment \cite{Agnese:2013rvf}. With its remarkable sensitivity, the \mjd~ is expected to resoundingly refute or confirm these results.  Here, we choose to remain agnostic about them and  assume that the DM is just at the border of what is allowed by the null results~\cite{Akerib:2010pv,Angle:2011th,Aprile:2012nq,Felizardo:2011uw} (see the black curves in Fig.~\ref{sensitivity} for current constraints). If the CoGeNT/DAMA/CDMS signals are confirmed, the \mjd~ will see even more events than we consider below.

If the \mjd~ indeed sees hundreds or thousands of DM events, the experiment will obviously try to determine the DM mass and cross section from this data. The accuracy of this determination, on very general grounds, is expected to depend on our knowledge of the ``beam", which in this case is provided by the local dark matter distribution. As we discuss below, its characteristics are uncertain. 
Two approaches to this uncertainty could be taken. One is to rely on models of the Galactic halo (analytical and/or numerical), the other is to try to extract \emph{both} the DM physics and astrophysics directly from the data. Previous studies examining the effect of astrophysical uncertainty on mass and cross section determination~\cite{Green:2007rb,Green:2008rd,Strigari:2009zb,Peter:2009ak,Pato:2010zk,McCabe:2010zh,Peter:2011eu,Pato:2011de,Kavanagh:2012nr,Pato:2012fw} involve some combinations of these approaches (though none deal with such light WIMPs).

Below, we highlight both approaches and then concentrate on the second one, which may become feasible with large statistics and hence the {\sc Demonstrator} may be the most appropriate place to test it.
 We will quantify how much simultaneously fitting for the DM properties and astrophysics degrades the determination of the DM mass. We will show that the culprit responsible for the degradation is a degeneracy that exists between the light WIMP mass and the DM velocity. 

To avoid any confusion, we stress from the outset that for the present purpose astrophysics obviously cannot be ``integrated out'', along the lines of~\cite{Fox:2010bz,Fox:2010bu}.  The power of the ``integrating out'' technique is that it allows mapping the results of one experiment into another in an astrophysics-free manner (see~\cite{Frandsen:2011gi}, \cite{Gondolo:2012rs} and \cite{HerreroGarcia:2012fu} for applications to recent experiments). For the purpose of measuring the DM properties, the astrophysics needs to instead be ``integrated in''. 

In the next section we review the basic theory of direct detection. In Sect.~\ref{fit} we discuss our fit methodology and mock-up of the \mjd.  In the first subsection of Sect.~\ref{sim} we present the results of fits to the mass, cross section and the local velocity dispersion, assuming the ``true'' (input) velocity distribution is Maxwell-Boltzmann.  
In subsection~\ref{VL2}, we examine to what extent light DM is sensitive to different forms of the velocity distribution by inputing the distribution from the high-resolution N-body simulation, Via Lactea II.  In Sect.~\ref{heavyWIMPs} we comment on the degeneracy at heavier masses ($m_{\chi}\sim100$ GeV) and in Sect.~\ref{streams} we briefly touch on the signatures of streams. We summarize our conclusions in Sect.~\ref{con}.

\section{Basics}
While the nature of WIMP-nucleus scattering is obviously unknown, in the vast majority of existing scenarios the scattering is mediated by a sufficiently heavy mediator particle (compared to the momentum transfer in the direct detection process). Upon integrating out the exchanged particle, in the usual effective field theory approach, one finds that all the physics of the DM-nucleus scattering is contained in a set of higher dimensional operators.  Several studies of the range of possibilities have been carried out~\cite{Fan:2010gt,Fitzpatrick:2012ix,Fitzpatrick:2012ib}.  A complete analysis would in principle examine how each of these operators is sensitive to astrophysical uncertainties in turn -- a daunting, but eventually necessary task. Here, mainly for clarity, we will vary the astrophysics while restricting ourselves to the simplest and most-studied interaction form: isospin-conserving, spin-independent scattering that depend on neither the incoming DM velocity nor the exchanged momentum.  Such a cross section could come from a scalar interaction $(\overline{q}q)(\overline{X}X)$ (such as from Higgs-exchange for a neutralino, e.g.~\cite{Jungman:1995df}) or a vector interaction $(\overline{q} \gamma_{\mu} q) (\overline{X} \gamma^{\mu} X)$ (arising from the exchange of a $Z'$ vector boson).
A complementary study was carried out in~\cite{McDermott:2011hx} where  the authors instead fixed the astrophysics and considered a 
set of particle physics choices.

To obtain the average differential rate per unit detector mass of a WIMP of mass $m_{X}$ scattering on a target nucleus of mass $m_{N}$, one convolves the cross section with the DM velocity distribution~(see~\cite{Jungman:1995df,Smith:1988kw} for reviews), 
\bea \frac{dR}{dE_{R}} &=& \frac{\rho_{\odot}}{m_{N}m_{X}}\left\langle v\frac{d \sigma}{dE_{R}} \right\rangle \\
&=& \frac{\rho_{\odot}}{m_{N}m_{X}} \int_{v_{min}(E_{R})}^{\infty} d^{3} v~ v f(\vec{v}+\vec{v}_{e}(t)) \frac{d\sigma}{dE_{R}}\nonumber, 
\eea
where $\mu_{N}$ is the DM-nucleus reduced mass, $\vec{v}_{e}(t)$ is the velocity of the laboratory observer with respect to the galactic rest frame, $f(v)$ is the local DM velocity distribution in the rest frame of the galaxy, and $\rho_{\odot}$ the local DM density. The quantity $v_{min} (E_{R})$ is the minimum DM velocity in the lab frame to produce a nuclear recoil of energy $E_{R}$; for elastic scattering, $v_{min} (E_{R}) = \sqrt{m_{N} E_{R}/2\mu_{N}^{2}}$. 

For spin- and momentum-independent interactions, the differential cross section can be written as
\be \frac{d \sigma}{d E_{R}} = \frac{m_{N}}{2\mu_{N}^{2}v^{2}} \sigma_{SI}^{N} F^{2} (E_{R},A),
\ee
where $F(E_{R},A)$ is the nuclear form factor. We use the Helm form factor, which can be found in~\cite{Lewin:1995rx}.  The spin-independent DM-nucleus cross section is 
\be
\sigma_{SI}^{N} =  \sigma_{n} \frac{\mu_{N}^{2}}{\mu_{n}^{2}} \frac{\left[ f_{p}Z + f_{n} (A-Z)\right]^{2}}{f_{n}^{2}},
\ee
where $\mu_{n}$ is the DM-nucleon reduced mass, and $f_{p}$ and $f_{n}$ are the DM couplings to protons and neutrons respectively.  Throughout, we will make the standard simplifying assumption of isospin-conserving scattering, $f_{p} = f_{n} = 1$.  With these assumptions, the scattering rate simplifies to
\be
\frac{dR}{dE_{R}} = \frac{\rho_{\odot} \sigma_{n}}{2\mu_{n}^{2}m_{X}}A^{2} F^{2}(E_{R},A) g(v_{min}),
\label{rate}
\ee
where $g(v_{min})$ is the mean inverse speed,
\be
g(v_{\rm{min}}) \equiv \int_{v_{\rm{min}}}^{\infty} \frac{f(\vec{v}+\vec{v}_{e}(t))}{v} d^{3}v .
\label{gvmin}
\ee

The astrophysical uncertainties in principle affect both the local density $\rho_{\odot}$ and the velocity distribution $f(v)$. In this letter, we adopt the standard fiducial value of the local DM density  $\rho_{\odot} = 0.3$~GeV/cm$^{3}$ and focus on velocities.  
With this framework, the unknown quantities are the DM mass $m_{X}$, scattering cross section $\sigma_{n}$ and $f(v)$ or, equivalently, $g(v_{min})$. 
\subsection{Velocity Distributions}
\label{sect:velocities}
While an order-of-magnitude estimate of the event rate can often be obtained by simply using the average velocity of a DM particle in the halo~\cite{Goodman:1984dc}, for a quantitative measurement of the DM properties the knowledge of the dark velocity distribution is required. A canonical framework has emerged in the field, in which the DM halo is taken to be an isothermal, Maxwell-Boltzmann (MB) distribution~\cite{Drukier:1986tm,Freese:1987wu}, with a cut-off corresponding to the escape speed. 
In the galactic rest-frame the distribution,
\begin{equation}
  f_{MB}(\vec{v}) = \left\{
     \begin{array}{lr}
       N  e^{-v^2/v_{0}^{2}}, & v < v_{esc}\\
       0, & v > v_{esc}
     \end{array}
   \right. ,
   \label{gauss_param}
\end{equation}
is fully specified by just two parameters: its dispersion $v_{0}$ and the local escape speed $v_{esc}$. The relevant velocity integral Eq.(\ref{gvmin}) and the normalization constant $N$ have a closed form expression (see Appendix). Assuming an idealized isothermal halo, with a $\rho\propto r^{-2}$ DM density profile, the dispersion $v_{0}$ could further be equated to the circular speed, $v_{e}$. The circular speed is observable and is measured to be $\langle v_{e}(t) \rangle = 230$ km/s~\cite{Bovy:2009dr,McMillan:2009yr}. This framework is being used by all experimental collaborations in reporting their results in terms of $m_{X}$ and $\sigma_{n}$.

We would like to get a sense of the impact of astrophysical uncertainties on these results. Even if one chooses to stick with the assumption of an idealized isothermal halo, an important uncertainty comes from the error on the circular speed, which is $\sigma_{v_{e}} \simeq 30$ km/s~\cite{Bovy:2009dr,McMillan:2009yr}. This error arises from the determination of the distance from the Sun to the Galactic center. Thus, at the minimum, one can consider varying the $v_{e}$ and $v_{0}$ in unison, within, say, a $2\sigma$ error bar (see Fig.~\ref{sensitivity} below) and, further, varying the escape speed, which, according to the recent RAVE survey, is in the region $ 498~{\rm km/s} < v_{esc} < 608~{\rm km/s}$ at $90 \%$ CL~\cite{Smith:2006ym}.

Of course, this approach is based on a long list of assumptions. It  neglects a possible DM core\footnote{The size of which for an idealized isothermal sphere is a free parameter.} and the gravitational effects of the baryons in the Galaxy. The isothermal assumption itself is {\it ad hoc}, and indeed detailed N-body simulations appear to be better fit by NFW profiles, the velocities of which show non-Maxwellian features, especially on the high end~\cite{Lisanti:2010qx}. The isotropy assumption for the local velocity distribution is also not borne out by the simulations, {\it cf.} \cite{Kuhlen:2009vh}. Even more distinctive for direct detection is the possibility of unequilibrated velocity structures coming from the recent tidal destruction of DM subhalos. These can include streams~\cite{Freese:2003na,Savage:2006qr}, as well as more spatially homogeneous tidal debris in the form of sheets and plumes~\cite{Lisanti:2011as,Kuhlen:2012fz}.  

In view of all this, the possibility of \emph{direct experimental} determination of the velocity distribution is of great interest. In our analysis, we will use the form in Eq.~(\ref{gauss_param}), fix $v_{esc}$ to 544 km/s, and vary $v_{e}$ around its fiducial value, 230 km/s. We stress that in this approach, the form of Eq.~(\ref{gauss_param}) is simply a parametrization of our ignorance, selected  mainly for clarity, and not a physical model. As such, one should not expect that $v_{0}$ be correlated with $v_{e}$. An alternative philosophy would be to use a model of the Galaxy to relate the two, as done in for example~\cite{Strigari:2009zb}.

We will further investigate, in Sect.~\ref{VL2}, if the \mjd, given massive statistics, would be able to tell a difference between the (averaged) numerical velocities and the isothermal model. Finally, we will comment on some implications from streams in Sect.~\ref{streams}.

\section{Experimental Details and Fit Setup}
\label{fit}

\begin{figure}[t] 
\begin{center}
\includegraphics[width=1\columnwidth]{
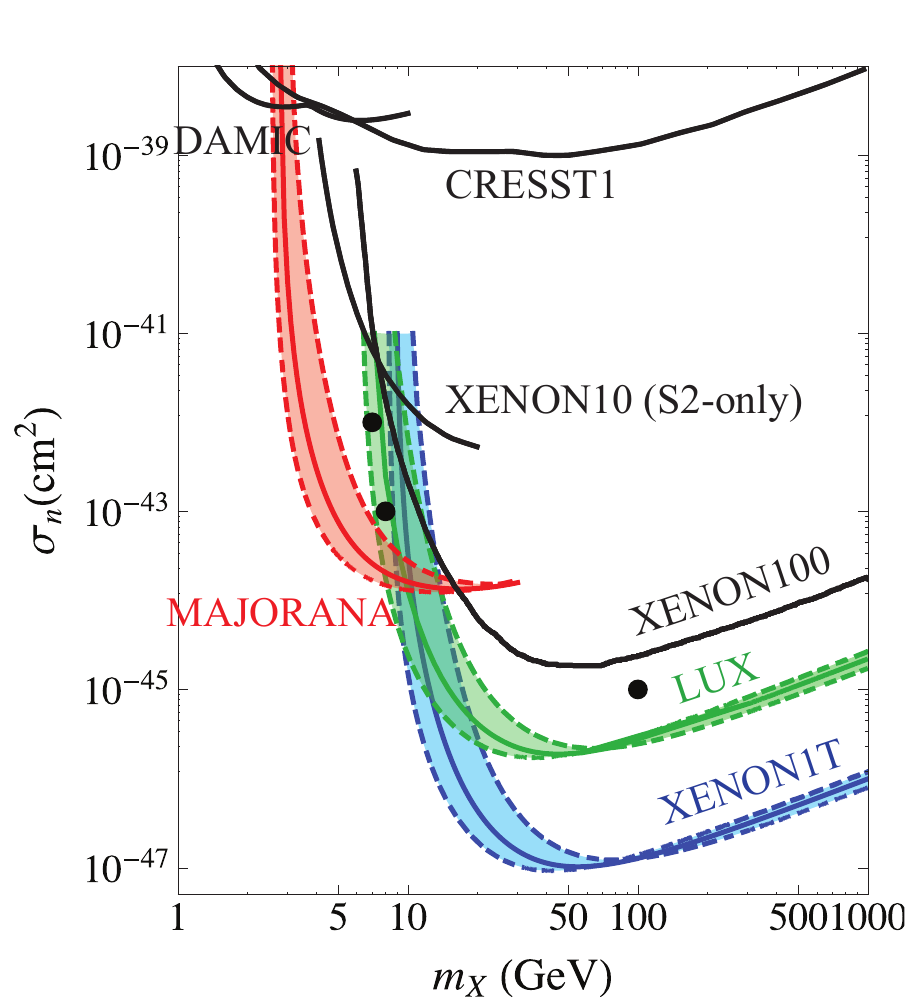}
\caption{Sensitivity reach of the upcoming and current experiments. The current constraints are shown with black curves: 2012 XENON100~\cite{Aprile:2012nq} along with the low-threshold analyses done by XENON10~\cite{Angle:2011th}, CRESST1~\cite{Altmann:2001ax} and DAMIC~\cite{Barreto:2011zu}. For the upcoming experiments, we use colored curves: solid assuming the velocity dispersion $v_{0}=230$ km/s and dashed for $v_{0}=170$ km/s and $v_{0}=290$ km/s. The circular speed $v_{e}$ is taken to be equal to $v_{0}$ here. For the {\sc Majorana Demonstrator} we assumed a 0.3 keV$_{ee}$ energy threshold and 100 kg-yr exposure. For LUX, we took a 100 kg-yr exposure with $E_{th} = 5$ keV, and for the XENON1T, a 2.2 ton-yr exposure and $E_{th} = 8.4$ keV  (see~\ref{heavyWIMPs} for more details). The black circles indicate points this paper examines in detail.}
\label{sensitivity}
\end{center}
\end{figure}

Though the {\sc Majorana Demonstrator}~\cite{Finnerty:2012dr} is primarily a neutrinoless double-beta decay experiment, the experimental setup is well-suited for DM direct detection as well~\cite{MajThesis}.  The {\sc Demonstrator} plans to achieve a very low threshold, in electron-equivalent units, 0.3-0.5 ${\rm keV_{ee}}$~\cite{MajThesis} (or equivalently $E_{NR} \approx 1.4-2.2$ keV in nuclear recoil). Such low thresholds are due to P-type contact detector technology~\cite{Giovanetti:2012ek}, also used by the CoGeNT experiment. Here we will take 0.3 ${\rm keV_{ee}}$ to compare with~\cite{MajThesis}.  The collaboration plans to have roughly 40 kg of Germanium detectors, deployed in three stages over the next two years. Here we will assume an exposure of 100 kg-yr~\cite{MajThesis}.  The {\sc Demonstrator} is now being constructed at the Sanford Underground Research Facility (SURF) in Lead, South Dakota~\cite{Finnerty:2012dr}. They plan to being data-taking in 2013, with the first set of two strings.

The DM sensitivity of \mjd~ is crucially affected by the background from cosmogenically activated tritium. This has been well studied in the thesis~\cite{MajThesis}. Assuming that a detector spends a total of 15 days at the surface, the average background rate contributed by decaying tritium is 0.03 counts/day/kg/keV~\cite{MajThesis}.  The tritium spectrum is well-known and extends to around 18 keV$_{ee}$. This background can of course fluctuate up and fake a WIMP signal. We will assume that this is the only relevant background in the signal region, and note that other possible backgrounds, including unknown ``surface events'' at CoGeNT(see for example~\cite{Kopp:2011yr}), need not impact the {\sc Demonstrator} in the same way, due to design differences.  

We also model the Xenon experiments, LUX and XENON1T, in a simplified manner. LUX is currently under construction at SURF while XENON1T will be installed at Gran Sasso National Laboratory in Italy, with planned data taking to begin in 2015. We mock up the XENON1T experiment by assuming a 2.2 ton-yr exposure with an energy window  $8.4 -44.6$  keV and a 40$\%$ acceptance~\cite{X1T}. We have checked that for heavy masses this mock-up agrees reasonably well with Fig. 2 of~\cite{Aprile:2012zx}. For the LUX experiment, we take 30000 kg-day exposure, the recoil energy window 5-30 keV and 45\% efficiency. These choices of the fixed low energy thresholds are conservative and facilitate comparisons with existing literature ({\it e.g.}, \cite{Peter:2011eu,Pato:2012fw}). The experiments themselves, however, may obtain sensitivity to recoil energies below the threshold used here, thereby extending their sensitivity to lower mass WIMPs. Thus in comparing with the XENON1T limits in~\cite{Aprile:2012zx} for example, our results reproduce well the exclusion sensitivity at high mass but are less stringent than what appears in~\cite{Aprile:2012zx}. We urge the experimental collaborations to improve on the analysis presented here.

\begin{figure*}[t] 
\begin{center}
\includegraphics[width=0.9\columnwidth]{
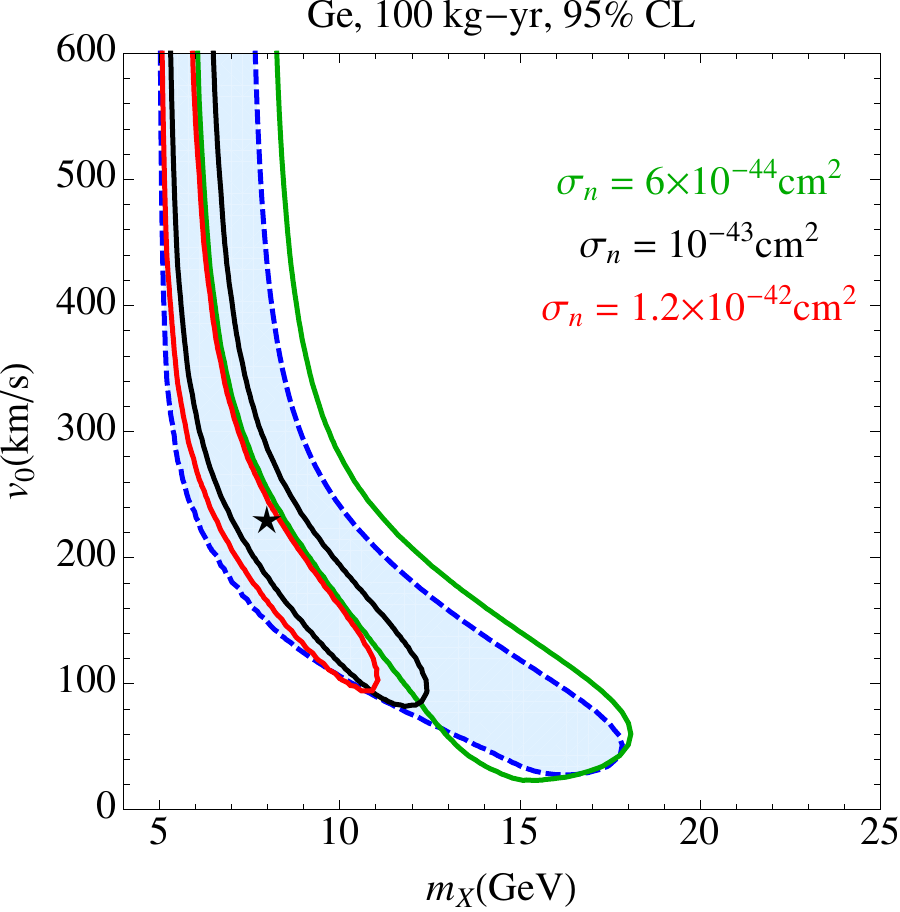}
\includegraphics[width=0.9\columnwidth]{
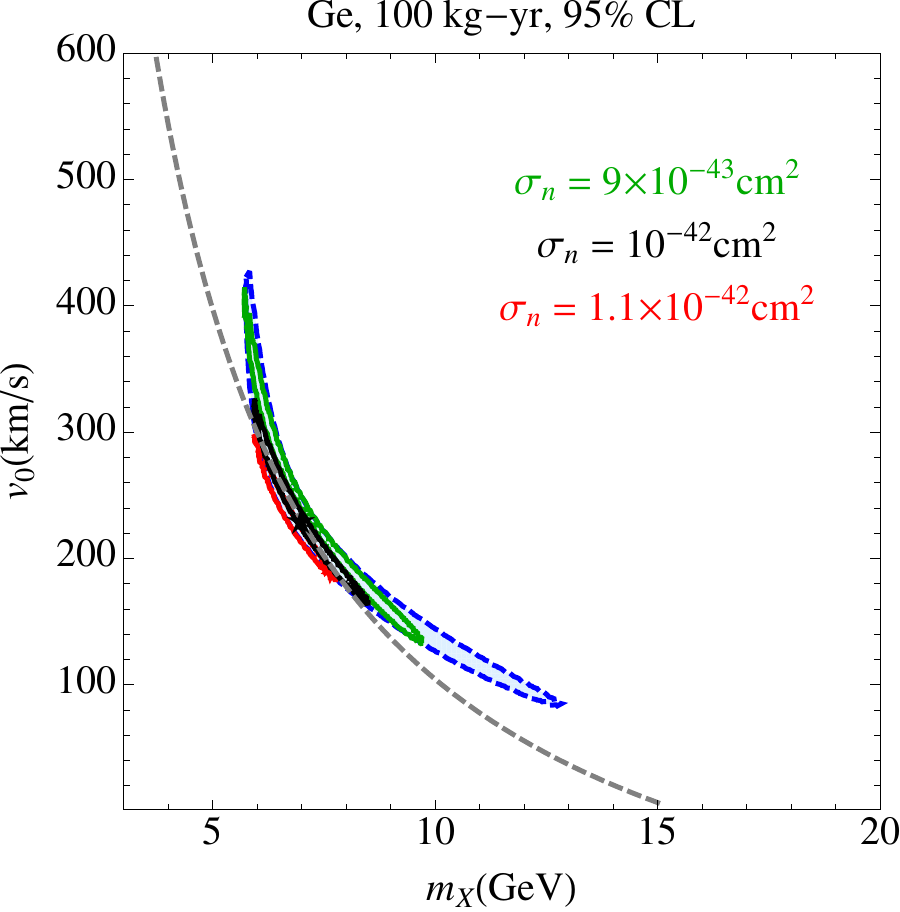}

\caption{Slices of the best-fit regions from the \mjd~ in $(m_{X},\sigma_{n}, v_{0})$ space. Light blue shows the cross section-marginalized regions, whereas the other contours are for fixed cross section as indicated in each plot.  The dashed gray curve represent the analytic estimate, Eq.~(\ref{erf}), which agrees well with the black contour. The input data is generated from a Maxwellian halo with with $v_{0} = v_{LSR} = 230$ km/s, with DM mass of 8 GeV and cross section $10^{-43}~\rm{cm}^{2}$ ({\it left})  and 7 GeV and cross section $10^{-42}~\rm{cm}^{2}$ ({\it right}).}
\label{degen}
\end{center}
\end{figure*}

For our statistical analysis, we bin events in $E_{R}$ (taking 0.9 keV bins for \mjd, and 3 keV bins for the Xenon experiments). 
The standard prescription for the likelihood of parameters is a product of Poisson distributions across the energy bins:
\be \mathcal{L}(\theta) = \prod_{i} \frac{ N(\theta,E_{i})^{N_{obs}(E_{i})}}{N_{obs}(E_{i})!} e^{-N (\theta,E_{i})}
\label{poisson}
\ee
where $N_{obs}(E_{i})$ is the observed number of events in the $i^{th}$ energy bin and $N (\theta,E_{i})$ is the number of expected events from parameters $\theta$ in the same bin. Given that at \mjd we will be examining cases with large numbers of events (from the signal plus the background), the Poisson distributions actually become Gaussians and the overall fit becomes well-described by a standard $\chi^{2}$ statistics. We have verified this explicitly.  

Moreover, since we are interested in the results obtained on average at a given experiment, we follow the method outlined in the Appendix of~\cite{deGouvea:1999wg}, and calculate the average $\chi^{2}$ as
\be \langle \chi^{2} \rangle = N_{dof} + \sum_{i} \frac{s_{i}^{2}}{b_{i}} \ee
where $b_{i}$ and $s_{i}$ are the background and signal expectation in the $i^{th}$ energy bin.  In the case of an exclusion, we calculate the limit by imposing $\langle \chi^{2} \rangle < \alpha_{CL} N_{dof}$, where $\alpha_{CL}$ depends on the desired confidence level. Throughout we will use 95$\%$ CL.  
When producing confidence regions from input signal spectra we calculate $\mathcal{L} \propto e^{-\langle \chi^{2} \rangle /2}$ at each parameter point, and then integrate the likelihood over the subspace of parameters that yield $95\%$ of the total likelihood. This method is useful for the non-Gaussian regions we obtain below. We return to Eq.~(\ref{poisson}) when we examine the case of XENON1T in Sect.~\ref{heavyWIMPs}.

\section{Simulation Results}
\label{sim}
\subsection{The Maxwell-Boltzmann Case with Light WIMPs}

As a first example, we illustrate the importance of the uncertainty on the dispersion in setting exclusion limits. We use the first framework outlined in Sect.~\ref{sect:velocities}, namely, vary $v_{e}$ and $v_{0}$ in lockstep, assuming the isothermal halo model. Null results at \mjd, LUX and XENON1T would result in exclusion curves shown in Fig.~\ref{sensitivity}. The solid curves assume the standard value $v_{0}=230$ km/s, while the bands show the result of varying $v_{0}$ within $\pm 60$ km/s.  For comparison, the existing exclusion curves are also shown (for fixed $v_{0}$). As is clear from the figure, the precise value of the velocity dispersion becomes particularly important for light WIMP masses, where the sensitivity of the experiments varies \emph{by an order of magnitude}. Physically, this is due to the fact that the events to which these experiments are sensitive all are caused by the high-velocity tail of the distribution.  The larger the value of $v_{0}$, the more stringent the bound for light masses. 

Let us now turn to our main simulation, of an actual WIMP detection, assuming the (unknown) parametrized velocity distribution of the form in Eq.~(\ref{gauss_param}). We will examine in detail two useful benchmark points: one with input data coming from a 8 GeV WIMP with scattering cross section $10^{-43}$ cm$^{2}$, and the other a 7 GeV WIMP with a $10^{-42}$ cm$^{2}$ cross section.  Both of these are just below the current XENON100 bounds~\cite{Aprile:2012nq}, as shown in Fig.~\ref{sensitivity}. At this mass and cross section, the \mjd~would see nearly 500 events in the 8 GeV case and around 3300 events in 7 GeV case.

\begin{figure*}[t] 
\begin{center}
\includegraphics[width=1\columnwidth]{
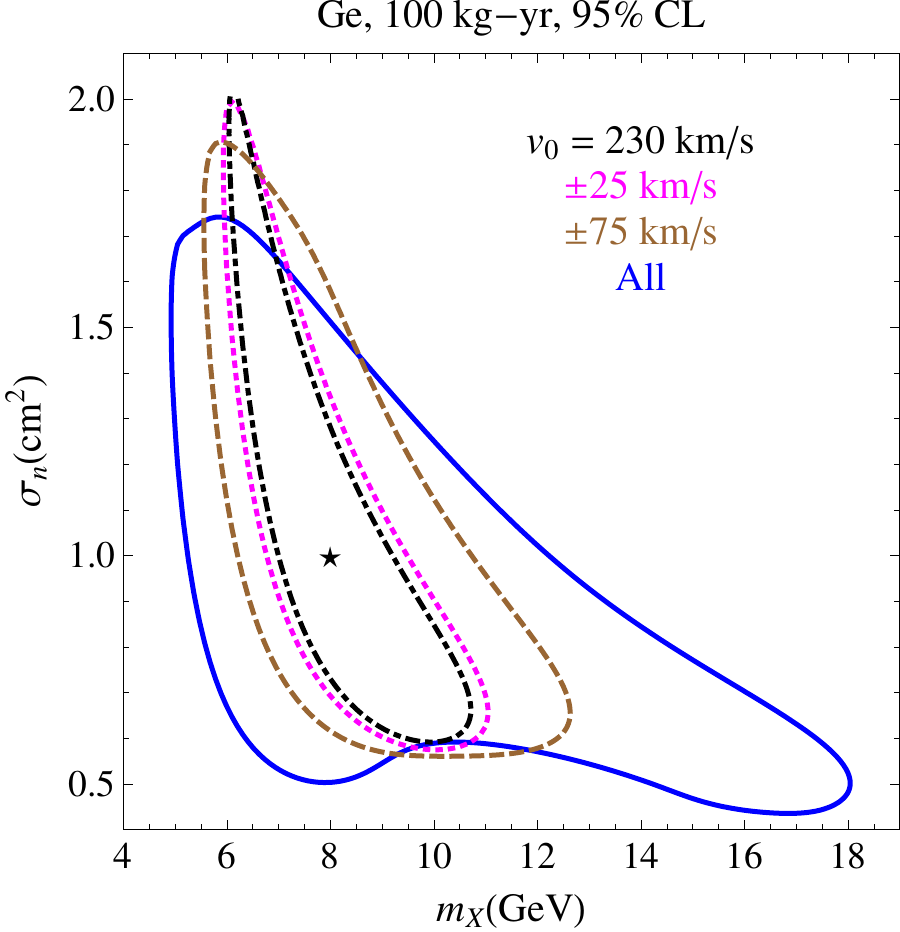}
\includegraphics[width=1\columnwidth]{
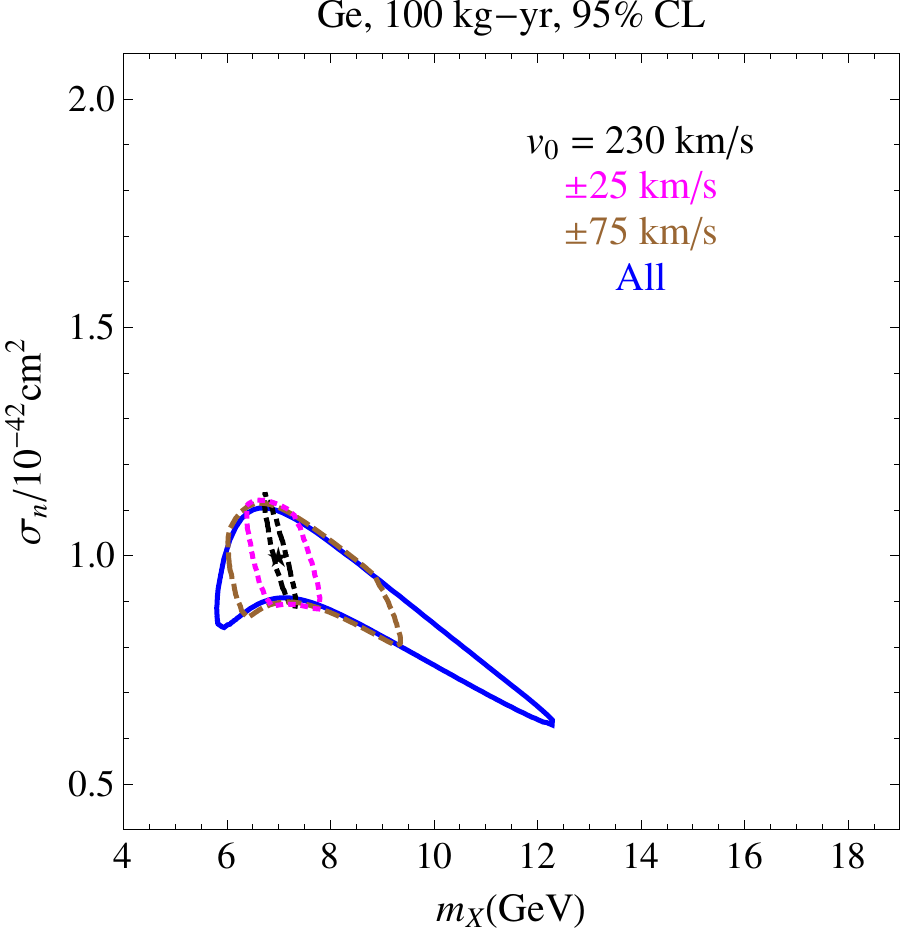}\\
\includegraphics[width=1\columnwidth]{
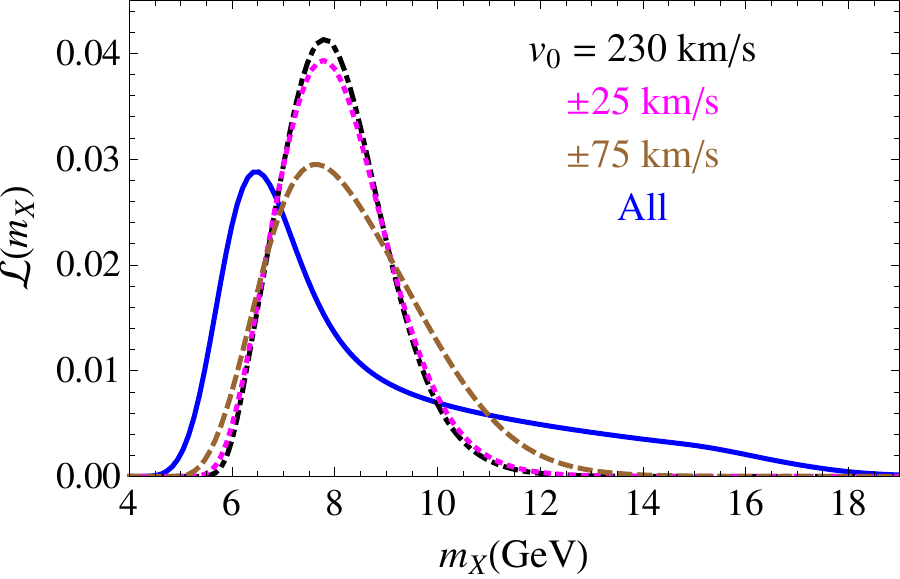}
\includegraphics[width=1\columnwidth]{
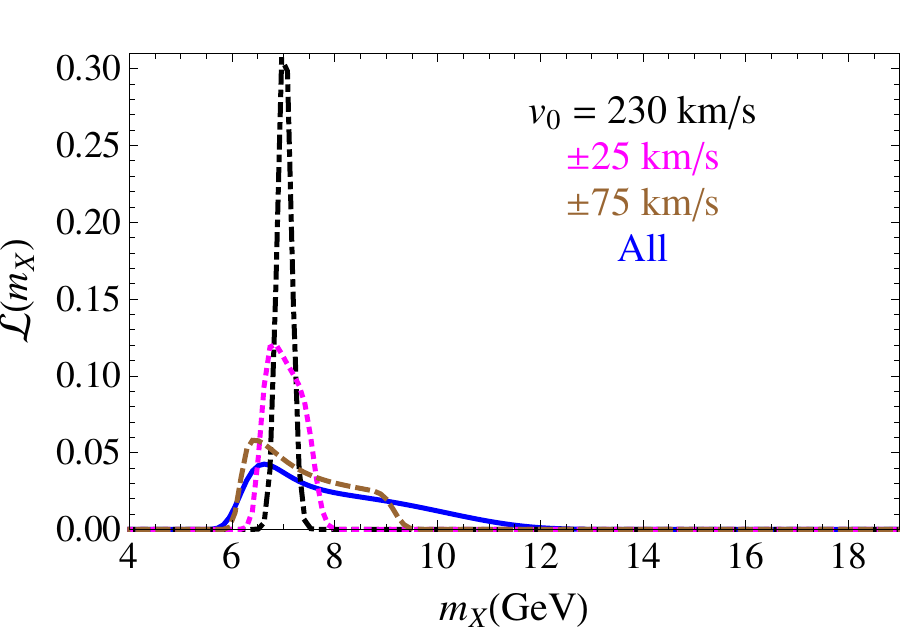}

\caption{Marginalized 95$\%$ CL regions for the {\sc Majorana Demonstrator} experiment with a 100 kg-yr exposure. In the upper panels we impose a top-hat prior on the dispersion, centered on the true value $230$ km/s with varying degrees of uncertainty $\sigma_{v}$. In the bottom row, we project this into the one-dimensional $m_{X}$ distribution. Results from a (8 GeV, $10^{-43}$ cm$^{2}$) and (7 GeV, $10^{-42}$ cm$^{2}$) input spectra are  depicted in the left and right columns respectively.  }
\label{marg}
\end{center}
\end{figure*}

For each of the two points, we generate ``data'', using $v_{e} =v_{0} = 230$ km/s and setting the escape speed to $v_{esc} = 544$ km/s. We then perform a fit on a three-dimensional grid of points, in the $(m_{X},\sigma_{n},v_{0})$ space. We keep  $v_{esc} = 544$ km/s and $v_{e} =230$ km/s both fixed in the fit. This is done primarily for the sake of clarity: to avoid confusion among many parameters. 


Let us first examine the 8 GeV case. In Fig.~\ref{degen} we show three slices of the full three-dimensional confidence regions at fixed cross sections. In addition to the input value (black), we also choose one lower value (in green) and one higher value (in red).  Marginalizing (integrating likelihood) over the cross section yields the light blue regions. 

We see a clear degeneracy between $m_{X}$ and $v_{0}$, which becomes particularly prominent in the case of larger statistics (in the right panel). Of course, one intuitively expects a degeneracy of this sort, since as one raises the fit mass parameter one must simultaneously lower the dispersion in order to produce a reasonable fit to the input spectrum (see for example, \cite{Green:2007rb,Green:2008rd,Kelso:2010sj,Farina:2011pw}). We would like, however, to give a quantitatively accurate description of it. To this end, let us specialize to the right panel in Fig.~\ref{degen} and consider a fixed cross section (the black region, for definiteness). Recall that for light WIMPs, $v_{min}$ is rather large (only high velocity WIMPs can produce enough recoil). This implies that the velocity integral $g(v_{min})$ (see the Appendix for the full expression) is well approximated by
\be g(v_{min}) \propto \left[ 1- {\rm erf}\left(\frac{v_{min}- v_{e}}{v_{0}}\right)\right],
\label{erf}
\ee
such that directions of constant $(v_{min}- v_{e})/v_{0}$ are degenerate. To translate this into a curve in the $m_{X}-v_{0}$ plane, we must fix $E_{R}$ or equivalently $v_{min}$. Taking 3 keV falls roughly in the middle of the signal spectrum. 

We display the analytic estimate based on Eq.~(\ref{erf}) as a dashed gray line in the right panel of Fig.~\ref{degen}.
It nicely reproduces the shape of the black region in Fig.~\ref{degen} \footnote{Note that this fit is better than the explicit parameterization given in~\cite{Green:2007rb,Green:2008rd}.}. Note that the contours at other cross sections can be understood with an analogous argument.

Notice that  the analytical contour bends away from the prediction once $v_{0}$ is greater than about 350 km/s. This is because of the effects of the finite the escape speed: for  $v_{0}\gtrsim350$ km and $v_{esc} = 544$ km/s , a significant part of the Maxwellian tail is cut off. (Eq.~(\ref{erf}) is obtained neglecting this cutoff.)  

Let us show how this degeneracy impacts the ability of the experiment to determine the mass and cross section of dark matter.  To do so, we marginalize out the $v_{0}$ dependence by imposing a top hat prior on the velocity dispersion centered on the true value 230 km/s, for a variety of uncertainties $\sigma_{v_{0}}$ ($\pm 25$ km/s, $\pm 75$ km/s and from zero to the escape speed).  The results for both examples are shown in the top two panels of Fig.~\ref{marg}, where we have also shown the regions coming from fixing the dispersion to 230 km/s (black).  The right panel is particularly instructive. We see that in the fixed astrophysics assumption one might expect an excellent measurement of the mass and cross section, but the measurement deteriorates significantly as the prior on $v_{0}$ get weaker and weaker. The long-tail extending to high masses results from the ability of high-mass/low-velocity dispersion cases to mimic the true data reasonably well.  The low-mass tail, which originates from the ability of low-mass/high-dispersion cases to match the true data, is considerably shorter.

Two comments are in order. First, notice that the degeneracy of $v_{0}$ with $\sigma_{n}$ is quite weak. This is because changing $v_{0}$ alters the signal spectrum in a way that depends on the target mass and does not reduce to a straightforward rescaling of the rates. It may be worth noting that this does not open the path of bringing the ``hint'' regions favored by different experiments \cite{Bernabei:2008yi,Aalseth:2010vx,Agnese:2013rvf} into closer agreement: the ``integrate out'' relations between the experiments hold in this case.  

Second, one may worry that relaxing the prior on $v_{0}$ appears to make some parts of the parameter space less favored. This happens because the contours show the most likely range of parameters, roughly speaking the $\Delta \chi^{2}$ contours relative to the best-fit point. If, as the prior is relaxed, a better fit is achieved somewhere, the $\Delta \chi^{2}$ contours may shift accordingly. This effect would be absent if we were to plot the goodness of fit contours (absolute $\chi^{2}$), instead of the parameter estimation contours.

\begin{figure}[t] 
\begin{center}
\includegraphics[width=0.9\columnwidth]{
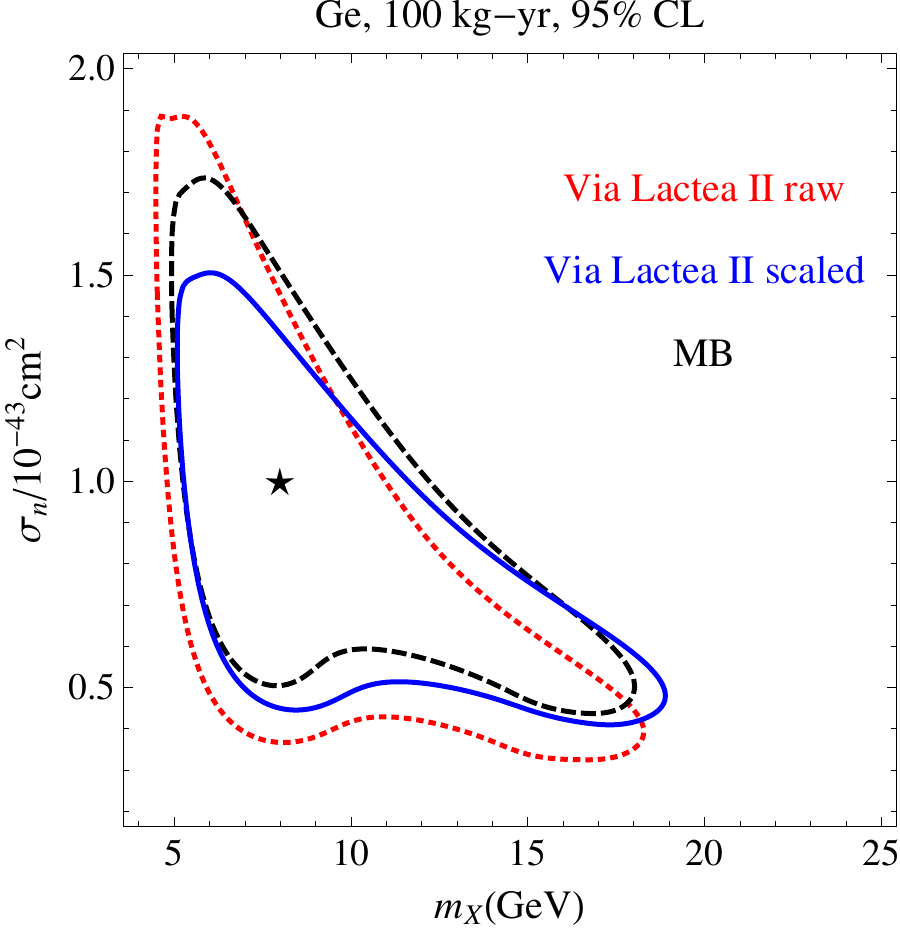}\\
\includegraphics[width=0.9\columnwidth]{
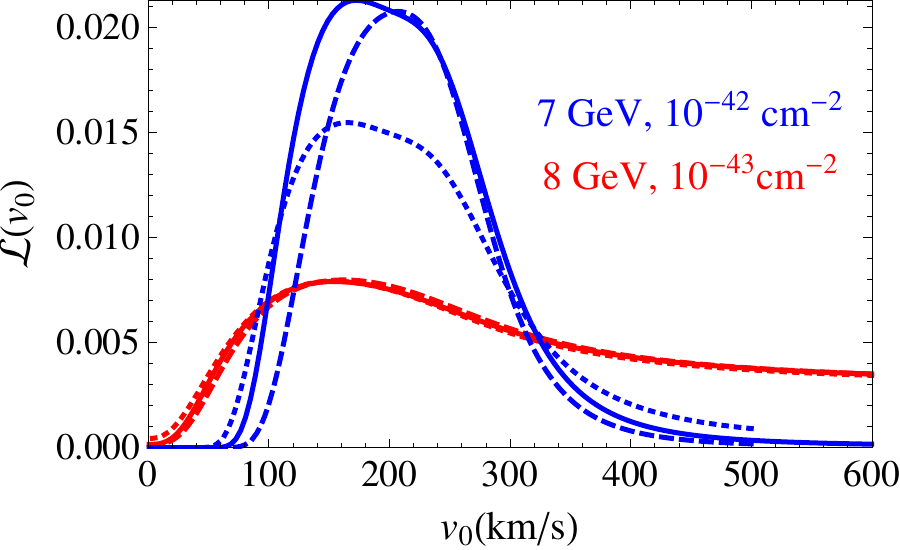}
\caption{Here we compare the CL regions when for the same ($m_{X} = 8$ GeV, $\sigma_{n} = 10^{-43}$ cm$^{2}$) example when coming from three different velocity distributions: Maxwell-Boltzmann (black, dashed), Via Lactea II raw distribution (red dotted) and Via Lactea II scaled to $v_{0} = 220$ km/s (blue, solid). {\it Bottom}: Likelihood functions of the velocity dispersion for all three distributions from above for both the 7 GeV and 8 GeV examples. Here solid, dashed, and dotted curves are MB, VL-II scaled and VL-II raw distributions. }
\label{VLcomp}
\end{center}
\end{figure}

Going one step further, we can marginalize out the cross section to see how well the WIMP mass, $m_{X}$, can be determined. The results are shown in the bottom panels in Fig.~\ref{marg}.  We first examine the 8 GeV case.  Interestingly, when the dispersion is varied, the experimental reconstruction prefers WIMP masses around 6 GeV. This is the result of projecting the ``sock-like'' region depicted in the panel above: the ``ankle'' contains many points along the $v_{0}$ direction, which when projected onto the $m_{X}$ axis, skews the likelihood toward low masses.

The same basic degeneracy and the concomitant skewed likelihood functions carries over for the 7 GeV case, despite the large increase in statistics. In the right panel of Fig.~\ref{marg} we repeat the above exercise for the example depicted in the right panel of Fig.~\ref{degen}. Again, this example includes a factor $\sim$7 increase in signal events, which is still insufficient to completely remove the degeneracy. In fact, the relative effect of the degeneracy is especially significant as the 1 $\sigma$ region expands from $m_{X} = 7^{+ 0.3}_{-0.2}$ GeV to $6.4^{+3.4}_{-0.1}$ GeV in going from fixed dispersion to complete uncertainty.

In summary, by inferring the astrophysics ($v_{0}$) from the data, we suffer a significant sensitivity loss to the light WIMP mass.

\subsection{Via Lactea II Velocity Distributions}
\label{VL2}

Do our findings depend on the particular choice of the mock spectrum? To check this, 
we repeat the analysis, but instead generate signal events from the high resolution N-body dark matter simulations borrowed from the Via Lactea II (VL-II) project~\cite{Diemand:2008in}. The effect of such distributions on direct detection phenomenology has been previously studied in~\cite{MarchRussell:2008dy,Kuhlen:2009vh} and constraints from existing experiments derived in~\cite{McCabe:2010zh}.  The effect of tidal debris in VL-II was first observed in~\cite{Lisanti:2011as} and subsequently applied to direct detection in~\cite{Kuhlen:2012fz,Cline:2012ei}, though here we focus on the full contents of the velocity distribution.

As we will show, the existence of the mass-dispersion degeneracy does not depend on the input spectrum coming from a Maxwell-Boltzmann form of the velocity distribution. Moreover, an experimentalist is unlikely to be able to exclude the Maxwellian hypothesis if the true halo is of a Via Lactea form.  Note that although the unscaled VL-II distribution is not a realistic approximation of the Milky Way halo, we include it here for comparison. The scaled VL-II distribution is scaled such that its dispersion is 220 km/s. 

As can be seen in Fig.~\ref{VLcomp} for the 8 GeV case, the same qualitative features of mass and dispersion mis-measurement persist as well, though differing somewhat in the quantitative details.  There we see the small effect induced from the raw VL-II distribution which favors low mass solutions compared to the other two distributions. An important observation is that the best-fit points have $\chi^{2}$ per degree of freedom that is very close to one, indicating that the isothermal fit is good. Interestingly, the values of $m_{X}$ and $\sigma_{n}$ obtained in the best-fit point are $\sim$15\% off from the original inputs. Nevertheless, since in an actual experiment the true (``input'') values are unknown, the fact that the distribution is non-Maxwellian will not be registered.

Lastly, we ask how well direct detection can determine the dispersion of the halo. One way of addressing this is to marginalize over the mass and cross section to determine the dispersion likelihood function, shown in the bottom panel of Fig.~\ref{VLcomp}. In the 8 GeV case (red curves) the dispersion likelihood function from {\sc Demonstrator} is extremely broad, with a poor ability to reconstruct the correct dispersion. Its best fit is peaked near 150 km/s, but with significant likelihood extending all the way to dispersions of 600 km/s. The high-statistics 7 GeV case fares a bit better with a best-fit value of 173 km/s for the MB and a best-fit of 207 km/s for the scaled VL-II distribution. Further, though there the 7 GeV starts to show visible separation of the three distributions, it is not sufficient for discriminating among velocity distributions.

\begin{figure}[t] 
\begin{center}
\includegraphics[width=0.9\columnwidth]{
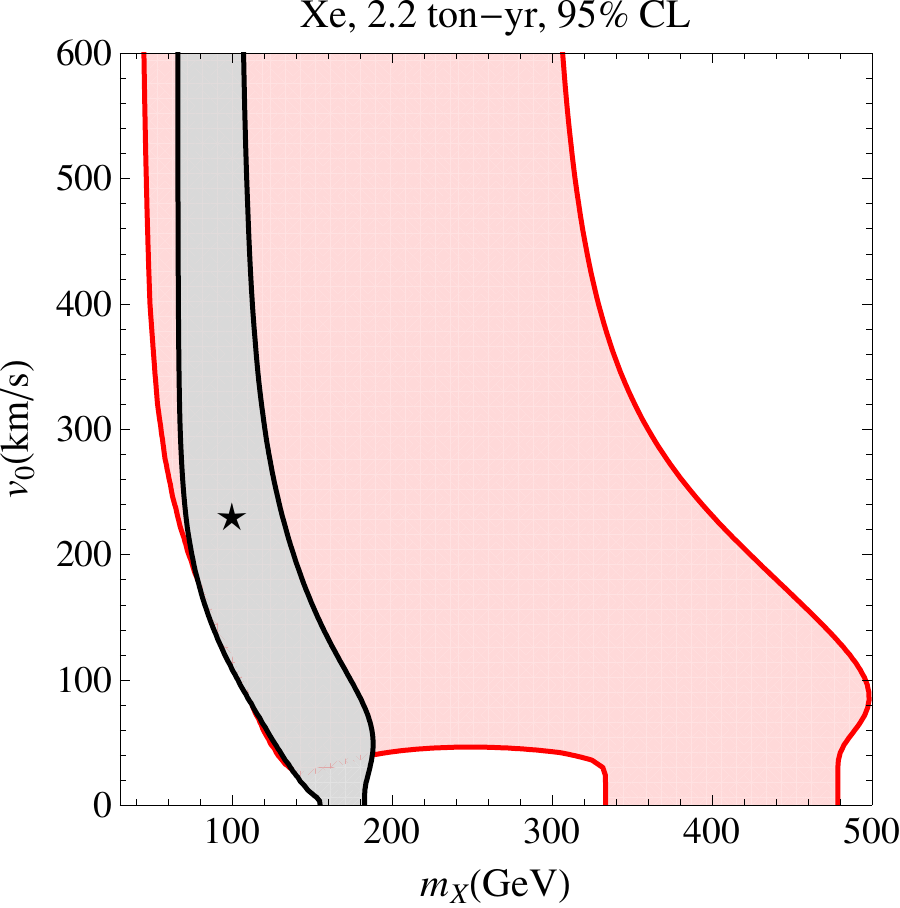}
\includegraphics[width=0.9\columnwidth]{
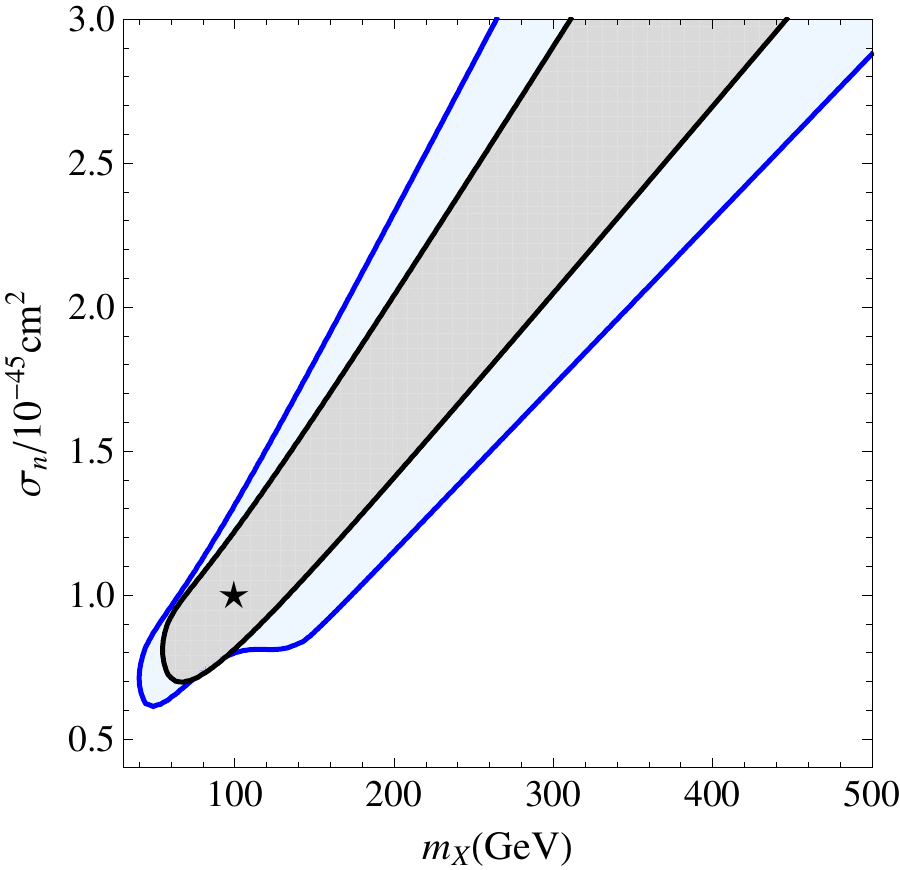}
\caption{Here the signal spectrum comes from a $v_{0} = 230$ km/s halo with WIMP mass 100 GeV and cross section $10^{-45}$ cm$^{2}$. In both figures, 68$\%$ (black) and 95$\%$ CL (red and blue) are displayed.  {\it Above}: One slice of the best-fit region with the cross section fixed to the true value, but floating $v_{0}$ and $m_{X}$. {\it Bottom}: The ensuing allowed regions coming from the canonical fit with fixed dispersion (black) and $v_{0}$ marginalized contour. }
\label{heavy}
\end{center}
\end{figure}

\subsection{Heavier WIMPs and XENON1T}
\label{heavyWIMPs}
In the above we have focused on the light WIMP case since the lowering of energy thresholds in upcoming experiments could easily lead to an immediate and huge signal. Significant statistics in the more canonical heavy WIMP case will require significantly larger experiments such as XENON1T~\cite{Aprile:2012zx}. 

We stress that the energy threshold we have chosen is conservative, and will hopefully be significantly lower given that XENON1T will have a larger light yield than XENON100~\cite{Aprile:2012zx}. Moreover, if XENON1T (or LUX) can lower their energy thresholds, they will have unprecedented sensitivity to a hitherto unexplored region of WIMP parameter space. Like the XENON10 and XENON100 experiments, XENON1T uses a combination of amplified charge signals (S2) and prompt scintillation light (S1) to discriminate signal from background.  Such low-thresholds are for example potentially achievable with the use of an S2-only analysis such as the one already performed by XENON10~\cite{Angle:2011th}. Alternatively, XENON1T may be able to use the ionized free electron signal as their primary energy estimator in order to obtain energy threshold approaching a few keV~\cite{Arisaka:2012ce}. Using the outer layers of the detector for self-shielding, the 3 tons of total liquid Xenon translate to 1.1 tons of fiducial detector mass.  

As an example, consider a 100 GeV WIMP with a 10$^{-45}$ cm$^{2}$ scattering cross section, which produces $\sim$100 events at our XENON1T mock-up.\footnote{Note that here we use a ``product of Poissons'' likelihood function as in Eq.~(\ref{poisson}) since the number of signal events in each bin can be small, though our results to not deviate significantly from a $\chi^{2}$ analysis.}  Though the relation describing the degeneracy at low masses (Eq.~\ref{erf}) no longer holds, we see in the top panel of Fig.~\ref{heavy} that a similar degeneracy persists at XENON1T. That the analytic description of the degeneracy does not hold here should not be surprising since $v_{min}$ is rather small in this case.

When the dispersion is marginalized out (bottom panel of Fig.~\ref{heavy}), we see that while the dispersion has an impact on the parameter reconstruction, it is less significant than in the light WIMP case.  In addition, the 95$\%$ CL contours (dashed) encounter a well known high mass degeneracy~\cite{Pato:2010zk}. This additional degeneracy appears because at sufficiently high WIMP mass $v_{min}$ no longer contains any information about $m_{X}$. Thus the only place in the differential scattering rate (Eq.~\ref{rate}) containing the WIMP mass is in the ratio $\sigma_{n}/m_{X}$ such that this combination forms an irreducible degenerate direction in parameter space. This example also offers poor prospects for dispersion determination, yielding at 1$\sigma$, $v_{0} = 144^{+294}_{-12}$ km/s. This skewed parameter determination is again due to the high mass/low dispersion models seen in the top panel of Fig.~\ref{heavy} that well mimic the input spectrum.

Lastly, one may wonder at what mass $m_{X}$ does the mass-degeneracy relation starts to take effect. In examining runs using our XENON1T setup, we find that by masses around 20 GeV the skewed $\sigma-m_{X}$ regions (such as Fig.~\ref{marg}) and ``sock-like" $m_{X}-v_{0}$ regions (such as Fig.~\ref{degen}) start to appear.  Thus even for more canonical ``high-threshold'' experiments the determination of the dispersion can be critical for precise DM mass measurements.

\subsection{Streams}
\label{streams}
Unbound streams of dark matter are generally expected in the hierarchical process of structure formation~\cite{Stiff:2001dq} as well as in non-standard halo models, such as the late-infall model~\cite{Copi:1999pw}.  The stream coming from the tidal disruption of the Sagittarius dwarf galaxy is perhaps the most well-studied such example~\cite{Freese:2003na,Freese:2003tt,Savage:2006qr,Natarajan:2011gz}, though N-body simulations predict a wealth of velocity substructure. One of their most dramatic signatures is in annual modulation searches, in which they can disrupt the prediction of a sinusoidal variation in time~\cite{Savage:2006qr,Natarajan:2011gz}. 

However, even at the level of the nuclear recoil spectrum, dispersionless streams produce rather distinctive phenomenology, with the stream contribution to $g(v_{min})$ appearing as a relatively sharp step-function. Generally, one expects such streams to form a subdominant contribution to the local density, $\xi \equiv \rho_{st}/\rho_{DM} < 1$, with the remainder of the local DM population described by an equilibrium distribution such as Maxwell-Boltzmann~\cite{Kamionkowski:2008vw}. Thus suppose when faced with a signal of DM, one is interested in excluding the MB only hypothesis. Then when fitting to a stream + MB signal with a MB distribution, we can roughly estimate an expected sensitivity to streams scaling as $N_{st} > \sqrt{N_{MB} + N_{bkg}}$, with $N_{st}, N_{MB}$, and $N_{bkg},$ being the number of events in the relevant energy bin(s) from the stream, MB halo, and background respectively. Thus, for irreducible backgrounds, the best sensitivity to streams occurs when the stream contributes to energy bins where the background dominates the MB signal. There one expects the ability to exclude the MB-only hypothesis when $N_{st} > \sqrt{N_{bkg}}$. We simulated spectra for the parameters of the Sagittarius stream, for which $v_{str} \approx 340$ km/s~\cite{Savage:2006qr}. With this scaling, in the case of the 7 GeV, $10^{-42}$ cm$^{2}$ example studied earlier, the \mjd~would allow 1$\sigma$ detection of the Sagittarius stream if its density is $\xi \gtrsim 6\%$.

Let us consider the other extreme case, in which the background is completely subdominant to the MB signal which has been recently studied in detail by~\cite{Natarajan:2011gz}.  There they have found that a 10 GeV WIMP capable of explaining the CoGeNT results, would have a systematic mass underestimate and cross section overestimate~\cite{Natarajan:2011gz}.  The authors of~\cite{Natarajan:2011gz}, find that a 5\% Sagittarius stream could be detectable at the $2\sigma$ level with a 10 kg-yr exposure on the C-4 CoGeNT upgrade.  We plan to return to this topic in the future, exploring the implications for the \mjd.

\section{Discussion and Conclusions}
\label{con}
At the outset of this paper we set out to ask a basic question that we may well be faced with shortly: If DM is a light WIMP with a large cross section, will we have sufficient information to infer both the particle physics and the local astrophysics of DM?  We have found that an accurate reconstruction of both the mass and the dispersion will be unlikely even when the signal consists of many thousands of events.  This is primarily due to a previously unexplored degeneracy between the WIMP mass and the local velocity dispersion. Indeed, the uncertainty on the average velocity of the local WIMP population is likely to be the largest factor inhibiting an accurate mass measurement.  

In addition, we have found that this conclusion holds if the signal spectra came from an isothermal halo or one described by the high resolution Via Lactea-II simulation.  We stress as well that despite these simulations including tidal debris which is made up of unequilibrated structure~\cite{Lisanti:2011as,Kuhlen:2012fz}, an experiment like the \mjd~will not be able to infer a deviation from the standard isothermal halo.  The phenomenologically more distinctive streams, however, could be well-identified with sufficient statistics.  Of course, improved sensitivity to streams and other tidal debris would come from the inclusion of modulation or directional information~(for recent work see~\cite{Bozorgnia:2011vc,Lee:2012pf,Alves:2012ay,Drukier:2012hj,Billard:2012qu,Morgan:2012sv}), which we have ignored in this first analysis.

In view of our findings, it is clear that dark matter mass determination could well hinge on our ability to better constrain the velocity distribution of the Milky Way by other means. Such an improvement could come about from improved analytic modeling, for example including self-consistent distributions that take into account the relevant astrophysical data along the lines of~\cite{Widrow:2008yg,Pato:2012fw}, as well as from more detailed and accurate numerical simulations of dark matter halos.  We plan to return to these questions in future work.

\acknowledgments
We would like to thank Elena Aprile, Michael Marino, Antonio J. Melgarejo, and Peter Sorenson for helpful correspondence. This work was supported by the LANL LDRD program.

\section*{Appendix: The Velocity Integral}

For completeness here we include a simple analytic expression for the MB, $f(v) \propto e^{-v^2/v_{0}^{2}}$ velocity integral relevant for direct detection~\cite{Smith:1988kw},\cite{Jungman:1995df},\cite{Savage:2006qr},\cite{McCabe:2010zh}. Note, however, that we do not include the unrealistic case in which $v_{obs} > v_{esc}$.  The velocity integral that determines scattering rates is defined as: 
\be
g(v_{\rm{min}}) \equiv \int_{v_{\rm{min}}}^{\infty} \frac{f({\bf v}+ {\bf v}_{obs}(t))}{v} d^{3}v.
\ee
To simply the expressions, we use the shorthand: $x_{e} \equiv v_{e}/v_{0}$, $x_{min} \equiv v_{min}/v_{0}$, and $x_{esc} \equiv v_{esc}/v_{0}$:

\begin{widetext}
\begin{equation*} 
g(v_{min}) = \frac{1}{2 N x_{e} v_{0}}\left[ {\rm erf} \left( x_{min}+ x_{e} \right)- {\rm erf} \left( x_{min} - x_{e}\right) - \frac{4 x_{e}}{\sqrt{\pi}}e^{-x_{esc}^2}\right], ~~x_{min} < |x_{esc} - x_{e}|
\end{equation*}

\begin{equation*} 
g(v_{min}) = \frac{1}{2 N x_{e} v_{0}}\left[ {\rm erf} \left( x_{esc} \right) + {\rm erf} \left(  x_{e}- x_{min}\right) - \frac{2}{\sqrt{\pi}}(x_{esc}+x_{e}- x_{min}) e^{-x_{esc}^2}\right], ~~  |x_{esc} - x_{e}| < x_{min} < x_{esc} + x_{e}
\end{equation*} 
\begin{equation*} 
g(v_{min}) = 0, ~~   x_{esc} + x_{e} < x_{min}
\end{equation*} 
\end{widetext}
where $N ={\rm erf}(x_{esc})- \frac{2x_{esc}}{\sqrt{\pi}} e^{-x_{esc}^{2}}.$

\bibliography{DMvel.bib}

\begin{thebibliography}{81}
\expandafter\ifx\csname natexlab\endcsname\relax\def\natexlab#1{#1}\fi
\expandafter\ifx\csname bibnamefont\endcsname\relax
  \def\bibnamefont#1{#1}\fi
\expandafter\ifx\csname bibfnamefont\endcsname\relax
  \def\bibfnamefont#1{#1}\fi
\expandafter\ifx\csname citenamefont\endcsname\relax
  \def\citenamefont#1{#1}\fi
\expandafter\ifx\csname url\endcsname\relax
  \def\url#1{\texttt{#1}}\fi
\expandafter\ifx\csname urlprefix\endcsname\relax\def\urlprefix{URL }\fi
\providecommand{\bibinfo}[2]{#2}
\providecommand{\eprint}[2][]{\url{#2}}

\bibitem[{\citenamefont{Goodman and Witten}(1985)}]{Goodman:1984dc}
\bibinfo{author}{\bibfnamefont{M.~W.} \bibnamefont{Goodman}} \bibnamefont{and}
  \bibinfo{author}{\bibfnamefont{E.}~\bibnamefont{Witten}},
  \bibinfo{journal}{Phys.Rev.} \textbf{\bibinfo{volume}{D31}},
  \bibinfo{pages}{3059} (\bibinfo{year}{1985}).

\bibitem[{\citenamefont{Marino}(2010)}]{MajThesis}
\bibinfo{author}{\bibfnamefont{M.~G.} \bibnamefont{Marino}},
  \emph{\bibinfo{title}{Dark matter physics with p-type point-contact germanium
  detectors: Extending the physics reach of the majorana experiment}}
  (\bibinfo{year}{2010}).

\bibitem[{\citenamefont{Finnerty et~al.}(2012)}]{Finnerty:2012dr}
\bibinfo{author}{\bibfnamefont{P.}~\bibnamefont{Finnerty}} \bibnamefont{et~al.}
  (\bibinfo{collaboration}{MAJORANA Collaboration}) (\bibinfo{year}{2012}),
  \eprint{1210.2678}.

\bibitem[{\citenamefont{Nussinov}(1985)}]{Nussinov:1985xr}
\bibinfo{author}{\bibfnamefont{S.}~\bibnamefont{Nussinov}},
  \bibinfo{journal}{Phys.Lett.} \textbf{\bibinfo{volume}{B165}},
  \bibinfo{pages}{55} (\bibinfo{year}{1985}).

\bibitem[{\citenamefont{Barr et~al.}(1990)\citenamefont{Barr, Chivukula, and
  Farhi}}]{Barr:1990ca}
\bibinfo{author}{\bibfnamefont{S.~M.} \bibnamefont{Barr}},
  \bibinfo{author}{\bibfnamefont{R.~S.} \bibnamefont{Chivukula}},
  \bibnamefont{and} \bibinfo{author}{\bibfnamefont{E.}~\bibnamefont{Farhi}},
  \bibinfo{journal}{Phys.Lett.} \textbf{\bibinfo{volume}{B241}},
  \bibinfo{pages}{387} (\bibinfo{year}{1990}).

\bibitem[{\citenamefont{Hooper et~al.}(2005)\citenamefont{Hooper,
  March-Russell, and West}}]{Hooper:2004dc}
\bibinfo{author}{\bibfnamefont{D.}~\bibnamefont{Hooper}},
  \bibinfo{author}{\bibfnamefont{J.}~\bibnamefont{March-Russell}},
  \bibnamefont{and} \bibinfo{author}{\bibfnamefont{S.~M.} \bibnamefont{West}},
  \bibinfo{journal}{Phys.Lett.} \textbf{\bibinfo{volume}{B605}},
  \bibinfo{pages}{228} (\bibinfo{year}{2005}), \eprint{hep-ph/0410114}.

\bibitem[{\citenamefont{Agashe and Servant}(2005)}]{Agashe:2004bm}
\bibinfo{author}{\bibfnamefont{K.}~\bibnamefont{Agashe}} \bibnamefont{and}
  \bibinfo{author}{\bibfnamefont{G.}~\bibnamefont{Servant}},
  \bibinfo{journal}{JCAP} \textbf{\bibinfo{volume}{0502}}, \bibinfo{pages}{002}
  (\bibinfo{year}{2005}), \eprint{hep-ph/0411254}.

\bibitem[{\citenamefont{Farrar and Zaharijas}(2006)}]{Farrar:2005zd}
\bibinfo{author}{\bibfnamefont{G.~R.} \bibnamefont{Farrar}} \bibnamefont{and}
  \bibinfo{author}{\bibfnamefont{G.}~\bibnamefont{Zaharijas}},
  \bibinfo{journal}{Phys.Rev.Lett.} \textbf{\bibinfo{volume}{96}},
  \bibinfo{pages}{041302} (\bibinfo{year}{2006}), \eprint{hep-ph/0510079}.

\bibitem[{\citenamefont{Gudnason et~al.}(2006)\citenamefont{Gudnason, Kouvaris,
  and Sannino}}]{Gudnason:2006ug}
\bibinfo{author}{\bibfnamefont{S.~B.} \bibnamefont{Gudnason}},
  \bibinfo{author}{\bibfnamefont{C.}~\bibnamefont{Kouvaris}}, \bibnamefont{and}
  \bibinfo{author}{\bibfnamefont{F.}~\bibnamefont{Sannino}},
  \bibinfo{journal}{Phys.Rev.} \textbf{\bibinfo{volume}{D73}},
  \bibinfo{pages}{115003} (\bibinfo{year}{2006}), \eprint{hep-ph/0603014}.

\bibitem[{\citenamefont{Kaplan et~al.}(2009)\citenamefont{Kaplan, Luty, and
  Zurek}}]{Kaplan:2009ag}
\bibinfo{author}{\bibfnamefont{D.~E.} \bibnamefont{Kaplan}},
  \bibinfo{author}{\bibfnamefont{M.~A.} \bibnamefont{Luty}}, \bibnamefont{and}
  \bibinfo{author}{\bibfnamefont{K.~M.} \bibnamefont{Zurek}},
  \bibinfo{journal}{Phys.Rev.} \textbf{\bibinfo{volume}{D79}},
  \bibinfo{pages}{115016} (\bibinfo{year}{2009}), \eprint{0901.4117}.

\bibitem[{\citenamefont{Tulin et~al.}(2012)\citenamefont{Tulin, Yu, and
  Zurek}}]{Tulin:2012re}
\bibinfo{author}{\bibfnamefont{S.}~\bibnamefont{Tulin}},
  \bibinfo{author}{\bibfnamefont{H.-B.} \bibnamefont{Yu}}, \bibnamefont{and}
  \bibinfo{author}{\bibfnamefont{K.~M.} \bibnamefont{Zurek}},
  \bibinfo{journal}{JCAP} \textbf{\bibinfo{volume}{1205}}, \bibinfo{pages}{013}
  (\bibinfo{year}{2012}), \eprint{1202.0283}.

\bibitem[{\citenamefont{Graesser et~al.}(2011)\citenamefont{Graesser,
  Shoemaker, and Vecchi}}]{Graesser:2011wi}
\bibinfo{author}{\bibfnamefont{M.~L.} \bibnamefont{Graesser}},
  \bibinfo{author}{\bibfnamefont{I.~M.} \bibnamefont{Shoemaker}},
  \bibnamefont{and} \bibinfo{author}{\bibfnamefont{L.}~\bibnamefont{Vecchi}},
  \bibinfo{journal}{JHEP} \textbf{\bibinfo{volume}{1110}}, \bibinfo{pages}{110}
  (\bibinfo{year}{2011}), \eprint{1103.2771}.

\bibitem[{\citenamefont{Goodman et~al.}(2011)\citenamefont{Goodman, Ibe,
  Rajaraman, Shepherd, Tait et~al.}}]{Goodman:2010yf}
\bibinfo{author}{\bibfnamefont{J.}~\bibnamefont{Goodman}},
  \bibinfo{author}{\bibfnamefont{M.}~\bibnamefont{Ibe}},
  \bibinfo{author}{\bibfnamefont{A.}~\bibnamefont{Rajaraman}},
  \bibinfo{author}{\bibfnamefont{W.}~\bibnamefont{Shepherd}},
  \bibinfo{author}{\bibfnamefont{T.~M.} \bibnamefont{Tait}},
  \bibnamefont{et~al.}, \bibinfo{journal}{Phys.Lett.}
  \textbf{\bibinfo{volume}{B695}}, \bibinfo{pages}{185} (\bibinfo{year}{2011}),
  \eprint{1005.1286}.

\bibitem[{\citenamefont{Friedland et~al.}(2012)\citenamefont{Friedland,
  Graesser, Shoemaker, and Vecchi}}]{Friedland:2011za}
\bibinfo{author}{\bibfnamefont{A.}~\bibnamefont{Friedland}},
  \bibinfo{author}{\bibfnamefont{M.~L.} \bibnamefont{Graesser}},
  \bibinfo{author}{\bibfnamefont{I.~M.} \bibnamefont{Shoemaker}},
  \bibnamefont{and} \bibinfo{author}{\bibfnamefont{L.}~\bibnamefont{Vecchi}},
  \bibinfo{journal}{Phys.Lett.} \textbf{\bibinfo{volume}{B714}},
  \bibinfo{pages}{267} (\bibinfo{year}{2012}), \eprint{1111.5331}.

\bibitem[{\citenamefont{Arkani-Hamed et~al.}(1998)\citenamefont{Arkani-Hamed,
  Dimopoulos, and Dvali}}]{ArkaniHamed:1998rs}
\bibinfo{author}{\bibfnamefont{N.}~\bibnamefont{Arkani-Hamed}},
  \bibinfo{author}{\bibfnamefont{S.}~\bibnamefont{Dimopoulos}},
  \bibnamefont{and} \bibinfo{author}{\bibfnamefont{G.}~\bibnamefont{Dvali}},
  \bibinfo{journal}{Phys.Lett.} \textbf{\bibinfo{volume}{B429}},
  \bibinfo{pages}{263} (\bibinfo{year}{1998}), \eprint{hep-ph/9803315}.

\bibitem[{\citenamefont{Abulencia et~al.}(2006)}]{Abulencia:2006kk}
\bibinfo{author}{\bibfnamefont{A.}~\bibnamefont{Abulencia}}
  \bibnamefont{et~al.} (\bibinfo{collaboration}{CDF Collaboration}),
  \bibinfo{journal}{Phys.Rev.Lett.} \textbf{\bibinfo{volume}{97}},
  \bibinfo{pages}{171802} (\bibinfo{year}{2006}), \eprint{hep-ex/0605101}.

\bibitem[{\citenamefont{Bernabei et~al.}(2008)}]{Bernabei:2008yi}
\bibinfo{author}{\bibfnamefont{R.}~\bibnamefont{Bernabei}} \bibnamefont{et~al.}
  (\bibinfo{collaboration}{DAMA Collaboration}), \bibinfo{journal}{Eur.Phys.J.}
  \textbf{\bibinfo{volume}{C56}}, \bibinfo{pages}{333} (\bibinfo{year}{2008}),
  \eprint{0804.2741}.

\bibitem[{\citenamefont{Aalseth et~al.}(2011)}]{Aalseth:2010vx}
\bibinfo{author}{\bibfnamefont{C.}~\bibnamefont{Aalseth}} \bibnamefont{et~al.}
  (\bibinfo{collaboration}{CoGeNT collaboration}),
  \bibinfo{journal}{Phys.Rev.Lett.} \textbf{\bibinfo{volume}{106}},
  \bibinfo{pages}{131301} (\bibinfo{year}{2011}), \eprint{1002.4703}.

\bibitem[{\citenamefont{Agnese et~al.}(2013)}]{Agnese:2013rvf}
\bibinfo{author}{\bibfnamefont{R.}~\bibnamefont{Agnese}} \bibnamefont{et~al.}
  (\bibinfo{collaboration}{CDMS Collaboration}),
  \bibinfo{journal}{Phys.Rev.Lett.}  (\bibinfo{year}{2013}),
  \eprint{1304.4279}.

\bibitem[{\citenamefont{Akerib et~al.}(2010)}]{Akerib:2010pv}
\bibinfo{author}{\bibfnamefont{D.}~\bibnamefont{Akerib}} \bibnamefont{et~al.}
  (\bibinfo{collaboration}{CDMS Collaboration}), \bibinfo{journal}{Phys.Rev.}
  \textbf{\bibinfo{volume}{D82}}, \bibinfo{pages}{122004}
  (\bibinfo{year}{2010}), \eprint{1010.4290}.

\bibitem[{\citenamefont{Angle et~al.}(2011)}]{Angle:2011th}
\bibinfo{author}{\bibfnamefont{J.}~\bibnamefont{Angle}} \bibnamefont{et~al.}
  (\bibinfo{collaboration}{XENON10 Collaboration}),
  \bibinfo{journal}{Phys.Rev.Lett.} \textbf{\bibinfo{volume}{107}},
  \bibinfo{pages}{051301} (\bibinfo{year}{2011}), \eprint{1104.3088}.

\bibitem[{\citenamefont{Aprile et~al.}(2012)}]{Aprile:2012nq}
\bibinfo{author}{\bibfnamefont{E.}~\bibnamefont{Aprile}} \bibnamefont{et~al.}
  (\bibinfo{collaboration}{XENON100 Collaboration}),
  \bibinfo{journal}{Phys.Rev.Lett.} \textbf{\bibinfo{volume}{109}},
  \bibinfo{pages}{181301} (\bibinfo{year}{2012}), \eprint{1207.5988}.

\bibitem[{\citenamefont{Felizardo et~al.}(2012)\citenamefont{Felizardo, Girard,
  Morlat, Fernandes, Ramos et~al.}}]{Felizardo:2011uw}
\bibinfo{author}{\bibfnamefont{M.}~\bibnamefont{Felizardo}},
  \bibinfo{author}{\bibfnamefont{T.}~\bibnamefont{Girard}},
  \bibinfo{author}{\bibfnamefont{T.}~\bibnamefont{Morlat}},
  \bibinfo{author}{\bibfnamefont{A.}~\bibnamefont{Fernandes}},
  \bibinfo{author}{\bibfnamefont{A.}~\bibnamefont{Ramos}},
  \bibnamefont{et~al.}, \bibinfo{journal}{Phys.Rev.Lett.}
  \textbf{\bibinfo{volume}{108}}, \bibinfo{pages}{201302}
  (\bibinfo{year}{2012}), \eprint{1106.3014}.

\bibitem[{\citenamefont{Green}(2007)}]{Green:2007rb}
\bibinfo{author}{\bibfnamefont{A.~M.} \bibnamefont{Green}},
  \bibinfo{journal}{JCAP} \textbf{\bibinfo{volume}{0708}}, \bibinfo{pages}{022}
  (\bibinfo{year}{2007}), \eprint{hep-ph/0703217}.

\bibitem[{\citenamefont{Green}(2008)}]{Green:2008rd}
\bibinfo{author}{\bibfnamefont{A.~M.} \bibnamefont{Green}},
  \bibinfo{journal}{JCAP} \textbf{\bibinfo{volume}{0807}}, \bibinfo{pages}{005}
  (\bibinfo{year}{2008}), \eprint{0805.1704}.

\bibitem[{\citenamefont{Strigari and Trotta}(2009)}]{Strigari:2009zb}
\bibinfo{author}{\bibfnamefont{L.~E.} \bibnamefont{Strigari}} \bibnamefont{and}
  \bibinfo{author}{\bibfnamefont{R.}~\bibnamefont{Trotta}},
  \bibinfo{journal}{JCAP} \textbf{\bibinfo{volume}{0911}}, \bibinfo{pages}{019}
  (\bibinfo{year}{2009}), \eprint{0906.5361}.

\bibitem[{\citenamefont{Peter}(2010)}]{Peter:2009ak}
\bibinfo{author}{\bibfnamefont{A.~H.} \bibnamefont{Peter}},
  \bibinfo{journal}{Phys.Rev.} \textbf{\bibinfo{volume}{D81}},
  \bibinfo{pages}{087301} (\bibinfo{year}{2010}), \eprint{0910.4765}.

\bibitem[{\citenamefont{Pato et~al.}(2011)\citenamefont{Pato, Baudis, Bertone,
  Ruiz~de Austri, Strigari et~al.}}]{Pato:2010zk}
\bibinfo{author}{\bibfnamefont{M.}~\bibnamefont{Pato}},
  \bibinfo{author}{\bibfnamefont{L.}~\bibnamefont{Baudis}},
  \bibinfo{author}{\bibfnamefont{G.}~\bibnamefont{Bertone}},
  \bibinfo{author}{\bibfnamefont{R.}~\bibnamefont{Ruiz~de Austri}},
  \bibinfo{author}{\bibfnamefont{L.~E.} \bibnamefont{Strigari}},
  \bibnamefont{et~al.}, \bibinfo{journal}{Phys.Rev.}
  \textbf{\bibinfo{volume}{D83}}, \bibinfo{pages}{083505}
  (\bibinfo{year}{2011}), \eprint{1012.3458}.

\bibitem[{\citenamefont{McCabe}(2010)}]{McCabe:2010zh}
\bibinfo{author}{\bibfnamefont{C.}~\bibnamefont{McCabe}},
  \bibinfo{journal}{Phys.Rev.} \textbf{\bibinfo{volume}{D82}},
  \bibinfo{pages}{023530} (\bibinfo{year}{2010}), \eprint{1005.0579}.

\bibitem[{\citenamefont{Peter}(2011)}]{Peter:2011eu}
\bibinfo{author}{\bibfnamefont{A.~H.} \bibnamefont{Peter}},
  \bibinfo{journal}{Phys.Rev.} \textbf{\bibinfo{volume}{D83}},
  \bibinfo{pages}{125029} (\bibinfo{year}{2011}), \eprint{1103.5145}.

\bibitem[{\citenamefont{Pato}(2011)}]{Pato:2011de}
\bibinfo{author}{\bibfnamefont{M.}~\bibnamefont{Pato}}, \bibinfo{journal}{JCAP}
  \textbf{\bibinfo{volume}{1110}}, \bibinfo{pages}{035} (\bibinfo{year}{2011}),
  \eprint{1106.0743}.

\bibitem[{\citenamefont{Kavanagh and Green}(2012)}]{Kavanagh:2012nr}
\bibinfo{author}{\bibfnamefont{B.~J.} \bibnamefont{Kavanagh}} \bibnamefont{and}
  \bibinfo{author}{\bibfnamefont{A.~M.} \bibnamefont{Green}},
  \bibinfo{journal}{Phys.Rev.} \textbf{\bibinfo{volume}{D86}},
  \bibinfo{pages}{065027} (\bibinfo{year}{2012}), \eprint{1207.2039}.

\bibitem[{\citenamefont{Pato et~al.}(2012)\citenamefont{Pato, Strigari, Trotta,
  and Bertone}}]{Pato:2012fw}
\bibinfo{author}{\bibfnamefont{M.}~\bibnamefont{Pato}},
  \bibinfo{author}{\bibfnamefont{L.~E.} \bibnamefont{Strigari}},
  \bibinfo{author}{\bibfnamefont{R.}~\bibnamefont{Trotta}}, \bibnamefont{and}
  \bibinfo{author}{\bibfnamefont{G.}~\bibnamefont{Bertone}}
  (\bibinfo{year}{2012}), \eprint{1211.7063}.

\bibitem[{\citenamefont{Fox et~al.}(2011{\natexlab{a}})\citenamefont{Fox, Liu,
  and Weiner}}]{Fox:2010bz}
\bibinfo{author}{\bibfnamefont{P.~J.} \bibnamefont{Fox}},
  \bibinfo{author}{\bibfnamefont{J.}~\bibnamefont{Liu}}, \bibnamefont{and}
  \bibinfo{author}{\bibfnamefont{N.}~\bibnamefont{Weiner}},
  \bibinfo{journal}{Phys.Rev.} \textbf{\bibinfo{volume}{D83}},
  \bibinfo{pages}{103514} (\bibinfo{year}{2011}{\natexlab{a}}),
  \eprint{1011.1915}.

\bibitem[{\citenamefont{Fox et~al.}(2011{\natexlab{b}})\citenamefont{Fox,
  Kribs, and Tait}}]{Fox:2010bu}
\bibinfo{author}{\bibfnamefont{P.~J.} \bibnamefont{Fox}},
  \bibinfo{author}{\bibfnamefont{G.~D.} \bibnamefont{Kribs}}, \bibnamefont{and}
  \bibinfo{author}{\bibfnamefont{T.~M.} \bibnamefont{Tait}},
  \bibinfo{journal}{Phys.Rev.} \textbf{\bibinfo{volume}{D83}},
  \bibinfo{pages}{034007} (\bibinfo{year}{2011}{\natexlab{b}}),
  \eprint{1011.1910}.

\bibitem[{\citenamefont{Frandsen et~al.}(2012)\citenamefont{Frandsen,
  Kahlhoefer, McCabe, Sarkar, and Schmidt-Hoberg}}]{Frandsen:2011gi}
\bibinfo{author}{\bibfnamefont{M.~T.} \bibnamefont{Frandsen}},
  \bibinfo{author}{\bibfnamefont{F.}~\bibnamefont{Kahlhoefer}},
  \bibinfo{author}{\bibfnamefont{C.}~\bibnamefont{McCabe}},
  \bibinfo{author}{\bibfnamefont{S.}~\bibnamefont{Sarkar}}, \bibnamefont{and}
  \bibinfo{author}{\bibfnamefont{K.}~\bibnamefont{Schmidt-Hoberg}},
  \bibinfo{journal}{JCAP} \textbf{\bibinfo{volume}{1201}}, \bibinfo{pages}{024}
  (\bibinfo{year}{2012}), \eprint{1111.0292}.

\bibitem[{\citenamefont{Gondolo and Gelmini}(2012)}]{Gondolo:2012rs}
\bibinfo{author}{\bibfnamefont{P.}~\bibnamefont{Gondolo}} \bibnamefont{and}
  \bibinfo{author}{\bibfnamefont{G.~B.} \bibnamefont{Gelmini}}
  (\bibinfo{year}{2012}), \eprint{1202.6359}.

\bibitem[{\citenamefont{Herrero-Garcia
  et~al.}(2012)\citenamefont{Herrero-Garcia, Schwetz, and
  Zupan}}]{HerreroGarcia:2012fu}
\bibinfo{author}{\bibfnamefont{J.}~\bibnamefont{Herrero-Garcia}},
  \bibinfo{author}{\bibfnamefont{T.}~\bibnamefont{Schwetz}}, \bibnamefont{and}
  \bibinfo{author}{\bibfnamefont{J.}~\bibnamefont{Zupan}},
  \bibinfo{journal}{Phys.Rev.Lett.} \textbf{\bibinfo{volume}{109}},
  \bibinfo{pages}{141301} (\bibinfo{year}{2012}), \eprint{1205.0134}.

\bibitem[{\citenamefont{Fan et~al.}(2010)\citenamefont{Fan, Reece, and
  Wang}}]{Fan:2010gt}
\bibinfo{author}{\bibfnamefont{J.}~\bibnamefont{Fan}},
  \bibinfo{author}{\bibfnamefont{M.}~\bibnamefont{Reece}}, \bibnamefont{and}
  \bibinfo{author}{\bibfnamefont{L.-T.} \bibnamefont{Wang}},
  \bibinfo{journal}{JCAP} \textbf{\bibinfo{volume}{1011}}, \bibinfo{pages}{042}
  (\bibinfo{year}{2010}), \eprint{1008.1591}.

\bibitem[{\citenamefont{Fitzpatrick
  et~al.}(2012{\natexlab{a}})\citenamefont{Fitzpatrick, Haxton, Katz, Lubbers,
  and Xu}}]{Fitzpatrick:2012ix}
\bibinfo{author}{\bibfnamefont{A.~L.} \bibnamefont{Fitzpatrick}},
  \bibinfo{author}{\bibfnamefont{W.}~\bibnamefont{Haxton}},
  \bibinfo{author}{\bibfnamefont{E.}~\bibnamefont{Katz}},
  \bibinfo{author}{\bibfnamefont{N.}~\bibnamefont{Lubbers}}, \bibnamefont{and}
  \bibinfo{author}{\bibfnamefont{Y.}~\bibnamefont{Xu}}
  (\bibinfo{year}{2012}{\natexlab{a}}), \eprint{1203.3542}.

\bibitem[{\citenamefont{Fitzpatrick
  et~al.}(2012{\natexlab{b}})\citenamefont{Fitzpatrick, Haxton, Katz, Lubbers,
  and Xu}}]{Fitzpatrick:2012ib}
\bibinfo{author}{\bibfnamefont{A.~L.} \bibnamefont{Fitzpatrick}},
  \bibinfo{author}{\bibfnamefont{W.}~\bibnamefont{Haxton}},
  \bibinfo{author}{\bibfnamefont{E.}~\bibnamefont{Katz}},
  \bibinfo{author}{\bibfnamefont{N.}~\bibnamefont{Lubbers}}, \bibnamefont{and}
  \bibinfo{author}{\bibfnamefont{Y.}~\bibnamefont{Xu}}
  (\bibinfo{year}{2012}{\natexlab{b}}), \eprint{1211.2818}.

\bibitem[{\citenamefont{Jungman et~al.}(1996)\citenamefont{Jungman,
  Kamionkowski, and Griest}}]{Jungman:1995df}
\bibinfo{author}{\bibfnamefont{G.}~\bibnamefont{Jungman}},
  \bibinfo{author}{\bibfnamefont{M.}~\bibnamefont{Kamionkowski}},
  \bibnamefont{and} \bibinfo{author}{\bibfnamefont{K.}~\bibnamefont{Griest}},
  \bibinfo{journal}{Phys.Rept.} \textbf{\bibinfo{volume}{267}},
  \bibinfo{pages}{195} (\bibinfo{year}{1996}), \eprint{hep-ph/9506380}.

\bibitem[{\citenamefont{McDermott et~al.}(2012)\citenamefont{McDermott, Yu, and
  Zurek}}]{McDermott:2011hx}
\bibinfo{author}{\bibfnamefont{S.~D.} \bibnamefont{McDermott}},
  \bibinfo{author}{\bibfnamefont{H.-B.} \bibnamefont{Yu}}, \bibnamefont{and}
  \bibinfo{author}{\bibfnamefont{K.~M.} \bibnamefont{Zurek}},
  \bibinfo{journal}{Phys.Rev.} \textbf{\bibinfo{volume}{D85}},
  \bibinfo{pages}{123507} (\bibinfo{year}{2012}), \eprint{1110.4281}.

\bibitem[{\citenamefont{Smith and Lewin}(1990)}]{Smith:1988kw}
\bibinfo{author}{\bibfnamefont{P.}~\bibnamefont{Smith}} \bibnamefont{and}
  \bibinfo{author}{\bibfnamefont{J.}~\bibnamefont{Lewin}},
  \bibinfo{journal}{Phys.Rept.} \textbf{\bibinfo{volume}{187}},
  \bibinfo{pages}{203} (\bibinfo{year}{1990}).

\bibitem[{\citenamefont{Lewin and Smith}(1996)}]{Lewin:1995rx}
\bibinfo{author}{\bibfnamefont{J.}~\bibnamefont{Lewin}} \bibnamefont{and}
  \bibinfo{author}{\bibfnamefont{P.}~\bibnamefont{Smith}},
  \bibinfo{journal}{Astropart.Phys.} \textbf{\bibinfo{volume}{6}},
  \bibinfo{pages}{87} (\bibinfo{year}{1996}).

\bibitem[{\citenamefont{Drukier et~al.}(1986)\citenamefont{Drukier, Freese, and
  Spergel}}]{Drukier:1986tm}
\bibinfo{author}{\bibfnamefont{A.}~\bibnamefont{Drukier}},
  \bibinfo{author}{\bibfnamefont{K.}~\bibnamefont{Freese}}, \bibnamefont{and}
  \bibinfo{author}{\bibfnamefont{D.}~\bibnamefont{Spergel}},
  \bibinfo{journal}{Phys.Rev.} \textbf{\bibinfo{volume}{D33}},
  \bibinfo{pages}{3495} (\bibinfo{year}{1986}).

\bibitem[{\citenamefont{Freese et~al.}(1988)\citenamefont{Freese, Frieman, and
  Gould}}]{Freese:1987wu}
\bibinfo{author}{\bibfnamefont{K.}~\bibnamefont{Freese}},
  \bibinfo{author}{\bibfnamefont{J.~A.} \bibnamefont{Frieman}},
  \bibnamefont{and} \bibinfo{author}{\bibfnamefont{A.}~\bibnamefont{Gould}},
  \bibinfo{journal}{Phys.Rev.} \textbf{\bibinfo{volume}{D37}},
  \bibinfo{pages}{3388} (\bibinfo{year}{1988}).

\bibitem[{\citenamefont{Bovy et~al.}(2009)\citenamefont{Bovy, Hogg, and
  Rix}}]{Bovy:2009dr}
\bibinfo{author}{\bibfnamefont{J.}~\bibnamefont{Bovy}},
  \bibinfo{author}{\bibfnamefont{D.~W.} \bibnamefont{Hogg}}, \bibnamefont{and}
  \bibinfo{author}{\bibfnamefont{H.-W.} \bibnamefont{Rix}},
  \bibinfo{journal}{Astrophys.J.} \textbf{\bibinfo{volume}{704}},
  \bibinfo{pages}{1704} (\bibinfo{year}{2009}), \eprint{0907.5423}.

\bibitem[{\citenamefont{McMillan and Binney}(2009)}]{McMillan:2009yr}
\bibinfo{author}{\bibfnamefont{P.~J.} \bibnamefont{McMillan}} \bibnamefont{and}
  \bibinfo{author}{\bibfnamefont{J.~J.} \bibnamefont{Binney}}
  (\bibinfo{year}{2009}), \eprint{0907.4685}.

\bibitem[{\citenamefont{Smith et~al.}(2007)\citenamefont{Smith, Ruchti, Helmi,
  Wyse, Fulbright et~al.}}]{Smith:2006ym}
\bibinfo{author}{\bibfnamefont{M.~C.} \bibnamefont{Smith}},
  \bibinfo{author}{\bibfnamefont{G.}~\bibnamefont{Ruchti}},
  \bibinfo{author}{\bibfnamefont{A.}~\bibnamefont{Helmi}},
  \bibinfo{author}{\bibfnamefont{R.}~\bibnamefont{Wyse}},
  \bibinfo{author}{\bibfnamefont{J.}~\bibnamefont{Fulbright}},
  \bibnamefont{et~al.}, \bibinfo{journal}{Mon.Not.Roy.Astron.Soc.}
  \textbf{\bibinfo{volume}{379}}, \bibinfo{pages}{755} (\bibinfo{year}{2007}),
  \eprint{astro-ph/0611671}.

\bibitem[{\citenamefont{Lisanti et~al.}(2011)\citenamefont{Lisanti, Strigari,
  Wacker, and Wechsler}}]{Lisanti:2010qx}
\bibinfo{author}{\bibfnamefont{M.}~\bibnamefont{Lisanti}},
  \bibinfo{author}{\bibfnamefont{L.~E.} \bibnamefont{Strigari}},
  \bibinfo{author}{\bibfnamefont{J.~G.} \bibnamefont{Wacker}},
  \bibnamefont{and} \bibinfo{author}{\bibfnamefont{R.~H.}
  \bibnamefont{Wechsler}}, \bibinfo{journal}{Phys.Rev.}
  \textbf{\bibinfo{volume}{D83}}, \bibinfo{pages}{023519}
  (\bibinfo{year}{2011}), \eprint{1010.4300}.

\bibitem[{\citenamefont{Kuhlen et~al.}(2010)\citenamefont{Kuhlen, Weiner,
  Diemand, Madau, Moore et~al.}}]{Kuhlen:2009vh}
\bibinfo{author}{\bibfnamefont{M.}~\bibnamefont{Kuhlen}},
  \bibinfo{author}{\bibfnamefont{N.}~\bibnamefont{Weiner}},
  \bibinfo{author}{\bibfnamefont{J.}~\bibnamefont{Diemand}},
  \bibinfo{author}{\bibfnamefont{P.}~\bibnamefont{Madau}},
  \bibinfo{author}{\bibfnamefont{B.}~\bibnamefont{Moore}},
  \bibnamefont{et~al.}, \bibinfo{journal}{JCAP}
  \textbf{\bibinfo{volume}{1002}}, \bibinfo{pages}{030} (\bibinfo{year}{2010}),
  \eprint{0912.2358}.

\bibitem[{\citenamefont{Freese et~al.}(2004)\citenamefont{Freese, Gondolo,
  Newberg, and Lewis}}]{Freese:2003na}
\bibinfo{author}{\bibfnamefont{K.}~\bibnamefont{Freese}},
  \bibinfo{author}{\bibfnamefont{P.}~\bibnamefont{Gondolo}},
  \bibinfo{author}{\bibfnamefont{H.~J.} \bibnamefont{Newberg}},
  \bibnamefont{and} \bibinfo{author}{\bibfnamefont{M.}~\bibnamefont{Lewis}},
  \bibinfo{journal}{Phys.Rev.Lett.} \textbf{\bibinfo{volume}{92}},
  \bibinfo{pages}{111301} (\bibinfo{year}{2004}), \eprint{astro-ph/0310334}.

\bibitem[{\citenamefont{Savage et~al.}(2006)\citenamefont{Savage, Freese, and
  Gondolo}}]{Savage:2006qr}
\bibinfo{author}{\bibfnamefont{C.}~\bibnamefont{Savage}},
  \bibinfo{author}{\bibfnamefont{K.}~\bibnamefont{Freese}}, \bibnamefont{and}
  \bibinfo{author}{\bibfnamefont{P.}~\bibnamefont{Gondolo}},
  \bibinfo{journal}{Phys.Rev.} \textbf{\bibinfo{volume}{D74}},
  \bibinfo{pages}{043531} (\bibinfo{year}{2006}), \eprint{astro-ph/0607121}.

\bibitem[{\citenamefont{Lisanti and Spergel}(2011)}]{Lisanti:2011as}
\bibinfo{author}{\bibfnamefont{M.}~\bibnamefont{Lisanti}} \bibnamefont{and}
  \bibinfo{author}{\bibfnamefont{D.~N.} \bibnamefont{Spergel}}
  (\bibinfo{year}{2011}), \eprint{1105.4166}.

\bibitem[{\citenamefont{Kuhlen et~al.}(2012)\citenamefont{Kuhlen, Lisanti, and
  Spergel}}]{Kuhlen:2012fz}
\bibinfo{author}{\bibfnamefont{M.}~\bibnamefont{Kuhlen}},
  \bibinfo{author}{\bibfnamefont{M.}~\bibnamefont{Lisanti}}, \bibnamefont{and}
  \bibinfo{author}{\bibfnamefont{D.~N.} \bibnamefont{Spergel}},
  \bibinfo{journal}{Phys.Rev.} \textbf{\bibinfo{volume}{D86}},
  \bibinfo{pages}{063505} (\bibinfo{year}{2012}), \eprint{1202.0007}.

\bibitem[{\citenamefont{Altmann et~al.}(2001)\citenamefont{Altmann, Angloher,
  Bruckmayer, Bucci, Cooper et~al.}}]{Altmann:2001ax}
\bibinfo{author}{\bibfnamefont{M.~F.} \bibnamefont{Altmann}},
  \bibinfo{author}{\bibfnamefont{G.}~\bibnamefont{Angloher}},
  \bibinfo{author}{\bibfnamefont{M.}~\bibnamefont{Bruckmayer}},
  \bibinfo{author}{\bibfnamefont{C.}~\bibnamefont{Bucci}},
  \bibinfo{author}{\bibfnamefont{S.}~\bibnamefont{Cooper}},
  \bibnamefont{et~al.} (\bibinfo{year}{2001}), \eprint{astro-ph/0106314}.

\bibitem[{\citenamefont{Barreto et~al.}(2012)}]{Barreto:2011zu}
\bibinfo{author}{\bibfnamefont{J.}~\bibnamefont{Barreto}} \bibnamefont{et~al.}
  (\bibinfo{collaboration}{DAMIC Collaboration}), \bibinfo{journal}{Phys.Lett.}
  \textbf{\bibinfo{volume}{B711}}, \bibinfo{pages}{264} (\bibinfo{year}{2012}),
  \eprint{1105.5191}.

\bibitem[{\citenamefont{Giovanetti et~al.}(2012)\citenamefont{Giovanetti,
  Aguayo, Avignone, Back, Barabash et~al.}}]{Giovanetti:2012ek}
\bibinfo{author}{\bibfnamefont{G.}~\bibnamefont{Giovanetti}},
  \bibinfo{author}{\bibfnamefont{E.}~\bibnamefont{Aguayo}},
  \bibinfo{author}{\bibfnamefont{F.}~\bibnamefont{Avignone}},
  \bibinfo{author}{\bibfnamefont{H.}~\bibnamefont{Back}},
  \bibinfo{author}{\bibfnamefont{A.}~\bibnamefont{Barabash}},
  \bibnamefont{et~al.}, \bibinfo{journal}{J.Phys.Conf.Ser.}
  \textbf{\bibinfo{volume}{375}}, \bibinfo{pages}{012014}
  (\bibinfo{year}{2012}).

\bibitem[{\citenamefont{Kopp et~al.}(2012)\citenamefont{Kopp, Schwetz, and
  Zupan}}]{Kopp:2011yr}
\bibinfo{author}{\bibfnamefont{J.}~\bibnamefont{Kopp}},
  \bibinfo{author}{\bibfnamefont{T.}~\bibnamefont{Schwetz}}, \bibnamefont{and}
  \bibinfo{author}{\bibfnamefont{J.}~\bibnamefont{Zupan}},
  \bibinfo{journal}{JCAP} \textbf{\bibinfo{volume}{1203}}, \bibinfo{pages}{001}
  (\bibinfo{year}{2012}), \eprint{1110.2721}.

\bibitem[{\citenamefont{Melgarejo}()}]{X1T}
\bibinfo{author}{\bibfnamefont{A.~J.} \bibnamefont{Melgarejo}},
  \bibinfo{note}{private communication}.

\bibitem[{\citenamefont{Aprile}(2012)}]{Aprile:2012zx}
\bibinfo{author}{\bibfnamefont{E.}~\bibnamefont{Aprile}}
  (\bibinfo{collaboration}{XENON1T collaboration}) (\bibinfo{year}{2012}),
  \eprint{1206.6288}.

\bibitem[{\citenamefont{de~Gouvea et~al.}(1999)\citenamefont{de~Gouvea,
  Friedland, and Murayama}}]{deGouvea:1999wg}
\bibinfo{author}{\bibfnamefont{A.}~\bibnamefont{de~Gouvea}},
  \bibinfo{author}{\bibfnamefont{A.}~\bibnamefont{Friedland}},
  \bibnamefont{and} \bibinfo{author}{\bibfnamefont{H.}~\bibnamefont{Murayama}},
  \bibinfo{journal}{Phys.Rev.} \textbf{\bibinfo{volume}{D60}},
  \bibinfo{pages}{093011} (\bibinfo{year}{1999}), \eprint{hep-ph/9904399}.

\bibitem[{\citenamefont{Kelso and Hooper}(2011)}]{Kelso:2010sj}
\bibinfo{author}{\bibfnamefont{C.}~\bibnamefont{Kelso}} \bibnamefont{and}
  \bibinfo{author}{\bibfnamefont{D.}~\bibnamefont{Hooper}},
  \bibinfo{journal}{JCAP} \textbf{\bibinfo{volume}{1102}}, \bibinfo{pages}{002}
  (\bibinfo{year}{2011}), \eprint{1011.3076}.

\bibitem[{\citenamefont{Farina et~al.}(2011)\citenamefont{Farina, Pappadopulo,
  Strumia, and Volansky}}]{Farina:2011pw}
\bibinfo{author}{\bibfnamefont{M.}~\bibnamefont{Farina}},
  \bibinfo{author}{\bibfnamefont{D.}~\bibnamefont{Pappadopulo}},
  \bibinfo{author}{\bibfnamefont{A.}~\bibnamefont{Strumia}}, \bibnamefont{and}
  \bibinfo{author}{\bibfnamefont{T.}~\bibnamefont{Volansky}},
  \bibinfo{journal}{JCAP} \textbf{\bibinfo{volume}{1111}}, \bibinfo{pages}{010}
  (\bibinfo{year}{2011}), \eprint{1107.0715}.

\bibitem[{\citenamefont{Diemand et~al.}(2008)\citenamefont{Diemand, Kuhlen,
  Madau, Zemp, Moore et~al.}}]{Diemand:2008in}
\bibinfo{author}{\bibfnamefont{J.}~\bibnamefont{Diemand}},
  \bibinfo{author}{\bibfnamefont{M.}~\bibnamefont{Kuhlen}},
  \bibinfo{author}{\bibfnamefont{P.}~\bibnamefont{Madau}},
  \bibinfo{author}{\bibfnamefont{M.}~\bibnamefont{Zemp}},
  \bibinfo{author}{\bibfnamefont{B.}~\bibnamefont{Moore}},
  \bibnamefont{et~al.}, \bibinfo{journal}{Nature}
  \textbf{\bibinfo{volume}{454}}, \bibinfo{pages}{735} (\bibinfo{year}{2008}),
  \eprint{0805.1244}.

\bibitem[{\citenamefont{March-Russell et~al.}(2009)\citenamefont{March-Russell,
  McCabe, and McCullough}}]{MarchRussell:2008dy}
\bibinfo{author}{\bibfnamefont{J.}~\bibnamefont{March-Russell}},
  \bibinfo{author}{\bibfnamefont{C.}~\bibnamefont{McCabe}}, \bibnamefont{and}
  \bibinfo{author}{\bibfnamefont{M.}~\bibnamefont{McCullough}},
  \bibinfo{journal}{JHEP} \textbf{\bibinfo{volume}{0905}}, \bibinfo{pages}{071}
  (\bibinfo{year}{2009}), \eprint{0812.1931}.

\bibitem[{\citenamefont{Cline et~al.}(2012)\citenamefont{Cline, Liu, and
  Xue}}]{Cline:2012ei}
\bibinfo{author}{\bibfnamefont{J.~M.} \bibnamefont{Cline}},
  \bibinfo{author}{\bibfnamefont{Z.}~\bibnamefont{Liu}}, \bibnamefont{and}
  \bibinfo{author}{\bibfnamefont{W.}~\bibnamefont{Xue}} (\bibinfo{year}{2012}),
  \eprint{1207.3039}.

\bibitem[{\citenamefont{Arisaka et~al.}(2012)\citenamefont{Arisaka, Beltrame,
  Ghag, Lung, and Scovell}}]{Arisaka:2012ce}
\bibinfo{author}{\bibfnamefont{K.}~\bibnamefont{Arisaka}},
  \bibinfo{author}{\bibfnamefont{P.}~\bibnamefont{Beltrame}},
  \bibinfo{author}{\bibfnamefont{C.}~\bibnamefont{Ghag}},
  \bibinfo{author}{\bibfnamefont{K.}~\bibnamefont{Lung}}, \bibnamefont{and}
  \bibinfo{author}{\bibfnamefont{P.~R.} \bibnamefont{Scovell}}
  (\bibinfo{year}{2012}), \eprint{1202.1924}.

\bibitem[{\citenamefont{Stiff et~al.}(2001)\citenamefont{Stiff, Widrow, and
  Frieman}}]{Stiff:2001dq}
\bibinfo{author}{\bibfnamefont{D.}~\bibnamefont{Stiff}},
  \bibinfo{author}{\bibfnamefont{L.~M.} \bibnamefont{Widrow}},
  \bibnamefont{and} \bibinfo{author}{\bibfnamefont{J.}~\bibnamefont{Frieman}},
  \bibinfo{journal}{Phys.Rev.} \textbf{\bibinfo{volume}{D64}},
  \bibinfo{pages}{083516} (\bibinfo{year}{2001}), \eprint{astro-ph/0106048}.

\bibitem[{\citenamefont{Copi et~al.}(1999)\citenamefont{Copi, Heo, and
  Krauss}}]{Copi:1999pw}
\bibinfo{author}{\bibfnamefont{C.~J.} \bibnamefont{Copi}},
  \bibinfo{author}{\bibfnamefont{J.}~\bibnamefont{Heo}}, \bibnamefont{and}
  \bibinfo{author}{\bibfnamefont{L.~M.} \bibnamefont{Krauss}},
  \bibinfo{journal}{Phys.Lett.} \textbf{\bibinfo{volume}{B461}},
  \bibinfo{pages}{43} (\bibinfo{year}{1999}), \eprint{hep-ph/9904499}.

\bibitem[{\citenamefont{Freese et~al.}(2005)\citenamefont{Freese, Gondolo, and
  Newberg}}]{Freese:2003tt}
\bibinfo{author}{\bibfnamefont{K.}~\bibnamefont{Freese}},
  \bibinfo{author}{\bibfnamefont{P.}~\bibnamefont{Gondolo}}, \bibnamefont{and}
  \bibinfo{author}{\bibfnamefont{H.~J.} \bibnamefont{Newberg}},
  \bibinfo{journal}{Phys.Rev.} \textbf{\bibinfo{volume}{D71}},
  \bibinfo{pages}{043516} (\bibinfo{year}{2005}), \eprint{astro-ph/0309279}.

\bibitem[{\citenamefont{Natarajan et~al.}(2011)\citenamefont{Natarajan, Savage,
  and Freese}}]{Natarajan:2011gz}
\bibinfo{author}{\bibfnamefont{A.}~\bibnamefont{Natarajan}},
  \bibinfo{author}{\bibfnamefont{C.}~\bibnamefont{Savage}}, \bibnamefont{and}
  \bibinfo{author}{\bibfnamefont{K.}~\bibnamefont{Freese}},
  \bibinfo{journal}{Phys.Rev.} \textbf{\bibinfo{volume}{D84}},
  \bibinfo{pages}{103005} (\bibinfo{year}{2011}), \eprint{1109.0014}.

\bibitem[{\citenamefont{Kamionkowski and
  Koushiappas}(2008)}]{Kamionkowski:2008vw}
\bibinfo{author}{\bibfnamefont{M.}~\bibnamefont{Kamionkowski}}
  \bibnamefont{and} \bibinfo{author}{\bibfnamefont{S.~M.}
  \bibnamefont{Koushiappas}}, \bibinfo{journal}{Phys.Rev.}
  \textbf{\bibinfo{volume}{D77}}, \bibinfo{pages}{103509}
  (\bibinfo{year}{2008}), \eprint{0801.3269}.

\bibitem[{\citenamefont{Bozorgnia et~al.}(2012)\citenamefont{Bozorgnia,
  Gelmini, and Gondolo}}]{Bozorgnia:2011vc}
\bibinfo{author}{\bibfnamefont{N.}~\bibnamefont{Bozorgnia}},
  \bibinfo{author}{\bibfnamefont{G.~B.} \bibnamefont{Gelmini}},
  \bibnamefont{and} \bibinfo{author}{\bibfnamefont{P.}~\bibnamefont{Gondolo}},
  \bibinfo{journal}{JCAP} \textbf{\bibinfo{volume}{1206}}, \bibinfo{pages}{037}
  (\bibinfo{year}{2012}), \eprint{1111.6361}.

\bibitem[{\citenamefont{Lee and Peter}(2012)}]{Lee:2012pf}
\bibinfo{author}{\bibfnamefont{S.~K.} \bibnamefont{Lee}} \bibnamefont{and}
  \bibinfo{author}{\bibfnamefont{A.~H.} \bibnamefont{Peter}},
  \bibinfo{journal}{JCAP} \textbf{\bibinfo{volume}{1204}}, \bibinfo{pages}{029}
  (\bibinfo{year}{2012}), \eprint{1202.5035}.

\bibitem[{\citenamefont{Alves et~al.}(2012)\citenamefont{Alves, Hedri, and
  Wacker}}]{Alves:2012ay}
\bibinfo{author}{\bibfnamefont{D.~S.} \bibnamefont{Alves}},
  \bibinfo{author}{\bibfnamefont{S.~E.} \bibnamefont{Hedri}}, \bibnamefont{and}
  \bibinfo{author}{\bibfnamefont{J.~G.} \bibnamefont{Wacker}}
  (\bibinfo{year}{2012}), \eprint{1204.5487}.

\bibitem[{\citenamefont{Drukier et~al.}(2012)\citenamefont{Drukier, Freese,
  Spergel, Cantor, Church et~al.}}]{Drukier:2012hj}
\bibinfo{author}{\bibfnamefont{A.}~\bibnamefont{Drukier}},
  \bibinfo{author}{\bibfnamefont{K.}~\bibnamefont{Freese}},
  \bibinfo{author}{\bibfnamefont{D.}~\bibnamefont{Spergel}},
  \bibinfo{author}{\bibfnamefont{C.}~\bibnamefont{Cantor}},
  \bibinfo{author}{\bibfnamefont{G.}~\bibnamefont{Church}},
  \bibnamefont{et~al.} (\bibinfo{year}{2012}), \eprint{1206.6809}.

\bibitem[{\citenamefont{Billard et~al.}(2012)\citenamefont{Billard, Riffard,
  Mayet, and Santos}}]{Billard:2012qu}
\bibinfo{author}{\bibfnamefont{J.}~\bibnamefont{Billard}},
  \bibinfo{author}{\bibfnamefont{Q.}~\bibnamefont{Riffard}},
  \bibinfo{author}{\bibfnamefont{F.}~\bibnamefont{Mayet}}, \bibnamefont{and}
  \bibinfo{author}{\bibfnamefont{D.}~\bibnamefont{Santos}}
  (\bibinfo{year}{2012}), \eprint{1207.1050}.

\bibitem[{\citenamefont{Morgan and Green}(2012)}]{Morgan:2012sv}
\bibinfo{author}{\bibfnamefont{B.}~\bibnamefont{Morgan}} \bibnamefont{and}
  \bibinfo{author}{\bibfnamefont{A.~M.} \bibnamefont{Green}},
  \bibinfo{journal}{Phys.Rev.} \textbf{\bibinfo{volume}{D86}},
  \bibinfo{pages}{083544} (\bibinfo{year}{2012}), \eprint{1208.4992}.

\bibitem[{\citenamefont{Widrow et~al.}(2008)\citenamefont{Widrow, Pym, and
  Dubinski}}]{Widrow:2008yg}
\bibinfo{author}{\bibfnamefont{L.~M.} \bibnamefont{Widrow}},
  \bibinfo{author}{\bibfnamefont{B.}~\bibnamefont{Pym}}, \bibnamefont{and}
  \bibinfo{author}{\bibfnamefont{J.}~\bibnamefont{Dubinski}},
  \bibinfo{journal}{Astrophys.J.} \textbf{\bibinfo{volume}{679}},
  \bibinfo{pages}{1239} (\bibinfo{year}{2008}), \eprint{0801.3414}.

\end{thebibliography}

\end{document}